\documentclass[onecolumn,floatfix, aps]{revtex4}
\usepackage{amsmath}
\usepackage{amssymb}
\usepackage{graphicx}
\usepackage{graphics}
\usepackage{subfigure}
\usepackage{epsfig}
\usepackage{amsfonts}
\usepackage[german, english]{babel}
\usepackage{ifthen}
\usepackage{pstricks}
\usepackage{xcolor}
\usepackage[active]{srcltx}
\DeclareGraphicsExtensions{.eps, .ps, .pdf}

\newcounter{fig}   

\begin{document}

\newcommand{\dd}{\mbox{d}}
\newcommand{\la}{\lambda}
\newcommand{\ta}{\theta}
\newcommand{\f}{\phi}
\newcommand{\vf}{\varphi}
\newcommand{\ka}{\kappa}
\newcommand{\al}{\alpha}
\newcommand{\ga}{\gamma}
\newcommand{\de}{\delta}
\newcommand{\si}{\sigma}
\newcommand{\bomega}{\mbox{\boldmath $\omega$}}
\newcommand{\bsi}{\mbox{\boldmath $\sigma$}}
\newcommand{\bchi}{\mbox{\boldmath $\chi$}}
\newcommand{\bal}{\mbox{\boldmath $\alpha$}}
\newcommand{\bpsi}{\mbox{\boldmath $\psi$}}
\newcommand{\brho}{\mbox{\boldmath $\varrho$}}
\newcommand{\beps}{\mbox{\boldmath $\varepsilon$}}
\newcommand{\bxi}{\mbox{\boldmath $\xi$}}
\newcommand{\bbeta}{\mbox{\boldmath $\beta$}}
\newcommand{\ee}{\end{equation}}
\newcommand{\eea}{\end{eqnarray}}
\newcommand{\be}{\begin{equation}}
\newcommand{\bea}{\begin{eqnarray}}
\newcommand{\ii}{\mbox{i}}
\newcommand{\e}{\mbox{e}}
\newcommand{\pa}{\partial}
\newcommand{\Om}{\Omega}
\newcommand{\vep}{\varepsilon}
\newcommand{\bfph}{{\bf \phi}}
\newcommand{\lm}{\lambda}

\newcommand{\alternative}[2]{ {\color{red} #1} {\color{blue} \textit{#2}} }
\newcommand{\corrected}[1]{}
\def\theequation{\arabic{equation}}
\newcommand{\re}[1]{(\ref{#1})}
\newcommand{\R}{{\rm I \hspace{-0.52ex} R}}
\newcommand{\N}{{\sf N\hspace*{-1.0ex}\rule{0.15ex}%
{1.3ex}\hspace*{1.0ex}}}
\newcommand{\Q}{{\sf Q\hspace*{-1.1ex}\rule{0.15ex}%
{1.5ex}\hspace*{1.1ex}}}
\newcommand{\C}{{\sf C\hspace*{-0.9ex}\rule{0.15ex}%
{1.3ex}\hspace*{0.9ex}}}
\newcommand{\eins}{1\hspace{-0.56ex}{\rm I}}

\title{Resonance structures in coupled two-component $\phi^4$ model}
\author{A.~Halavanau$^{1}$, T.~Romanczukiewicz$^{2}$ and Ya.~Shnir$^{1,3}$}

\affiliation{$^{1}$Department of Theoretical Physics and Astrophysics, BSU, Minsk, Belarus\\
$^{2}$Institute of Physics, Jagiellonian University, Krakow, Poland\\
$^{3}$Institute of Physics, Carl von Ossietzky University Oldenburg, Germany}

\begin{abstract}
We present a numerical study of the process of the kink-antikink collisions in the coupled
one-dimensional two-component $\phi^4$ model. Our results reveal two different soliton solutions
which represent double kink configuration and kink-non-topological soliton (lump) bound state.
Collision of these solitons leads to very reach resonance
structure which is related to reversible energy exchange between the kinks, non-topological
solitons and the internal vibrational modes. Various channels of the collisions are discussed,
it is shown there is a new type of self-similar fractal structure which appears in the collisions of
the relativistic kinks, there the width of the resonance windows increases with the increase of the impact velocity.
An analytical approximation scheme is discussed in the
limit of the perturbative coupling between the sectors. Considering the spectrum of
linear fluctuations around the solitons we found that the double kink configuration is unstable if the
coupling constant between the sectors is negative.
\end{abstract}

\maketitle

%\pacs{11.10.Lm, 11.27.+d}

%\medskip

%%%%%%%%%%%%%%%%%%%%%%%%%%%%%%%%%%%%%%%%%%%%%%%%%%%%%%%%%%%%%%%%%%
\section{Introduction}
%%%%%%%%%%%%%%%%%%%%%%%%%%%%%%%%%%%%%%%%%%%%%%%%%%%%%%%%%%%%%%%%%%
Since the early 1960s, the soliton solutions in the non-linear field theories have been intensively
studied in various frameworks.
It is evident these spatially localized non-perturbative configurations play
important role in a wide variety of physical systems. The study of the interaction between the
solitons and their dynamical properties has attracted a lot of attention in many different contexts.

Simplest example of the topological solitons in one dimension is the class of the kink ($K$) solutions
which appears in the model with a potential with two or more degenerated minima.
Double well potential
corresponds to the nonintegrable $\phi^4$ model. Further extentions of this model are possible,
for example one can consider domain walls obtained by embedding of the 1+1-dimensional $\phi^4$-kink into
higher dimensions.
This model has a number of applications in condensed matter physics \cite{Solitons},
field theory \cite{MantonSutcliffe,Vachaspati:2006zz} and cosmology \cite{Vilenkin}.
Probably one of the most fascinating developments in this direction is the idea that
one may regard our universe
as a domain wall \cite{Akama:1982jy,Rubakov2003}
and Big Bang is associated with the collision of the domain walls \cite{Khoury:2001wf}.

Further, in various branches of physics more complex systems involving several coupled scalar fields arise
(see, e.q. \cite{Rajaraman:1978kd,Bazeia:1995en,Malomed}). Properties of the solitons in these extended models
are rather different from being a trivial extension of the
single-component one, in particular the model may support existence of non-topological solitons \cite{Rajaraman:1978kd}
or it can be related with application of the supersymmetry in nonrelativistic quantum mechanics \cite{Rodrigues}.
However, relatively little is known about the dynamics of the multicomponent configurations in the extended non-integrable
models.

Dynamical properties of the usual $\phi^4$ kinks, the processes of their scattering, radiation and annihilation
have already been discussed in a number of papers, see e.g.
\cite{Makhankov:1978rg,Moshir:1981ja,Peyrard:1984qn,Campbell:1983xu,Anninos:1991un,Manton:1996ex,Belova:1997bq,Googman:2005}.
Surprisely, it was discovered numerically that there is dynamical fractal structure in the process of
the kink-antikink ($K\bar K$) collision \cite{Peyrard:1984qn,Campbell:1983xu,Anninos:1991un,Googman:2005}.
Depending on the impact velocity, the collision may produce various results,
for some range of values of the impact velocity the $K\bar K$ pair form an oscillating
$n$-bouncing window, after collision the pair reflects back to some finite separation and returns to collide $(n-1)$ times more
before escaping to infinity. Also the process of production of kink-antikink pairs in the collision of
particle-like states is chaotic \cite{Romanczukiewicz:2010eg}.
Another interesting phenomenon occurs when the kink configuration is under the influence of an incident wave \cite{Radiative},
it was found that the $\phi^4$ kink starts to accelerate towards the incoming wave although the effect depends on the structure
of the potential.

The explanation of the appearance of the resonant scattering windows in the $K\bar K$ collision is related to the energy
exchange between the translational collective mode of the solitons and an internal vibrational mode of the kink
\cite{Peyrard:1984qn,Campbell:1983xu,Anninos:1991un,Googman:2005}. Another mechanism works in the $\phi^6$ model where
the resonance windows appear due to energy transfer between collective bound states trapped by the $K\bar K$ pair \cite{Dorey:2011yw}.
Evidently, in the extended multicomponent model coupling to the second component may strongly affect the process of the resonant
energy transfer.

A natural extension of the simple model $\phi^4$ model in 1+1 dimension is related to
consideration of systems with two or more component scalar fields.
Several models of that type have been investigated in recent years, some of them are related
to consideration of the complex scalar field, a deformation of the linear O(2)-sigma model known
as MSTB-model \cite{Montonen:1976,Sarkar:1976vr,Hawrylak:1984rc} or with consideration of
a two-component non-linear interaction model
\cite{Rajaraman:1978kd,Bazeia:1995en,Bazeia:1997zp,AlonsoIzquierdo:2008jc,Romanczukiewicz:2008hi}. In particular,
this model supports existence of solutions
with one of the component having the kink structure and the second component
being a non-topological soliton.
Another situation occurs when the system of coupled scalar fields belongs to the bosonic sector of a
supersymmetric system (Wess-Zumino model with polynomic superpotential and two Majorana spinor fields)
\cite{AlonsoIzquierdo:2000mj}, or with application of supersymmetric methods \cite{Rodrigues} and
related reduction of the second-order field equations to the system of
corresponding first-order Bogomol'nyi equations \cite{de Souza Dutra:2006pu,de Souza Dutra:2007jh}.

The aim of the present paper is to analyse the process of the $K\bar K$ resonant bouncing scattering
in the two-component coupled $\phi^4$ model. We assume that both sectors are two replica of the
usual $\phi^4$ model, the fields are coupled through the minimal quadratic term
and the kink-antikink collisions are taking place in the first sector. Evidently the fine mechanism of the reversible
energy transfer in the bouncing of the kink will be modified in such a model because both the vacuum structure and
the spectrum of linear oscillations around the solitons will be different from the standard one-component $\phi^4$ theory.

This present work is organized in the following way. In Section II we overview the spectrum of soliton solutions
of the coupled two-component $\phi^4$ model and discuss the analytical perturbative
approximation to these configurations in both sectors. In section III we present our numerical results of  the
various kink-antikink ($K\bar K$) collisions and discuss the resonance structures we observed.
In section IV we describe the spectral structure of linear pertubations on the background of the soliton solutions
of this model and discuss the possible mechanisms of the resonance energy transfer between the internal modes and the kinks.
We present our conclusions in section V.

\section{Two-component coupled $\phi^4$ model}

We consider the system of two coupled copies of the $\phi^4$ fields
\be  \label{Lag}
L = \frac{1}{2}\partial_\mu \phi_1 \partial^\mu \phi_1 + \frac{1}{2}\partial_\mu \phi_2 \partial^\mu \phi_2
- \frac{1}{2}\left(\phi_1^2 -1\right)^2 - \frac{1}{2}\left(\phi_2^2 - a^2\right)^2 + \kappa \phi_1^2\phi_2^2
\, .
\ee
where $a$ is associated with the mass of the field $\phi_2$ and $\kappa$ is the coupling parameter \footnote{Note that usual
classical rescaling of the 2-component model in one spacial dimension does not allow us to absorb all the constants into rescaled
scalar fields and couplings.}. Thus, the potential of the model is
\be
\label{pot}
U(\phi_1,\phi_2)=\frac{1}{2}\left(\phi_1^2 -1\right)^2 + \frac{1}{2}\left(\phi_2^2 - a^2\right)^2 -
\kappa \phi_1^2\phi_2^2
\ee
where $\kappa$ is the coupling constant. Hereafter, for the sake of simplicity we set $a=1$.

Let us consider the vacuum manifold of the model which is defined by the conditions
\be
\phi_{1,t}=\phi_{2,t}=\phi_{1,x}=\phi_{2,x} = 0,\; \qquad U(\phi_1,\phi_2) = \min,
\ee
that is
\begin{subequations}
\begin{align}
      2\phi_1(\phi_1^2-1-\kappa\phi_2^2) &= 0,\\
      2\phi_2(\phi_2^2-1-\kappa\phi_1^2) &= 0.
\end{align}
\end{subequations}
Evidently, the structure of the vacuum depends on the value of the coupling constant $\kappa$.
In the case when $-1<\kappa<1$ there are nine stationary points of the potential \re{pot}:
\begin{itemize}
   \item one local maximum $\phi_1=\phi_2=0$\quad ($U_{max}=1$),
   \item four saddle points $\phi_1=0, \phi_2=\pm 1$ or $\phi_1=\pm 1, \phi_2=0$ \quad ($U_{saddle} = \frac{1}{2}$),
   \item four minima $\phi_1=\pm 1/\sqrt{1-\kappa},\; \phi_2=\pm 1/\sqrt{1-\kappa}$ \quad ($U_{min} = -\frac{\kappa}{1-\kappa}$).
\end{itemize}

If $\kappa>1$ the system becomes unbounded from below, that is it is unstable, so we shall not consider this case.
In the limiting case $\kappa=-1$
the minima of the potential \re{pot} and its  saddle points are degenerated while in the regime $\kappa<-1$
the minima appear at the four points $\phi_1=0, \phi_2=\pm 1$ or $\phi_1=\pm 1, \phi_2=0$,
which previously, when $-1 < \kappa <1$,  were associated with the saddle points.
Correspondingly, the former minima become the saddle points in this case, see Fig~\ref{fig:1}.
\begin{figure}

\centering
%\begin{center}
\includegraphics[width=12cm,angle=0]{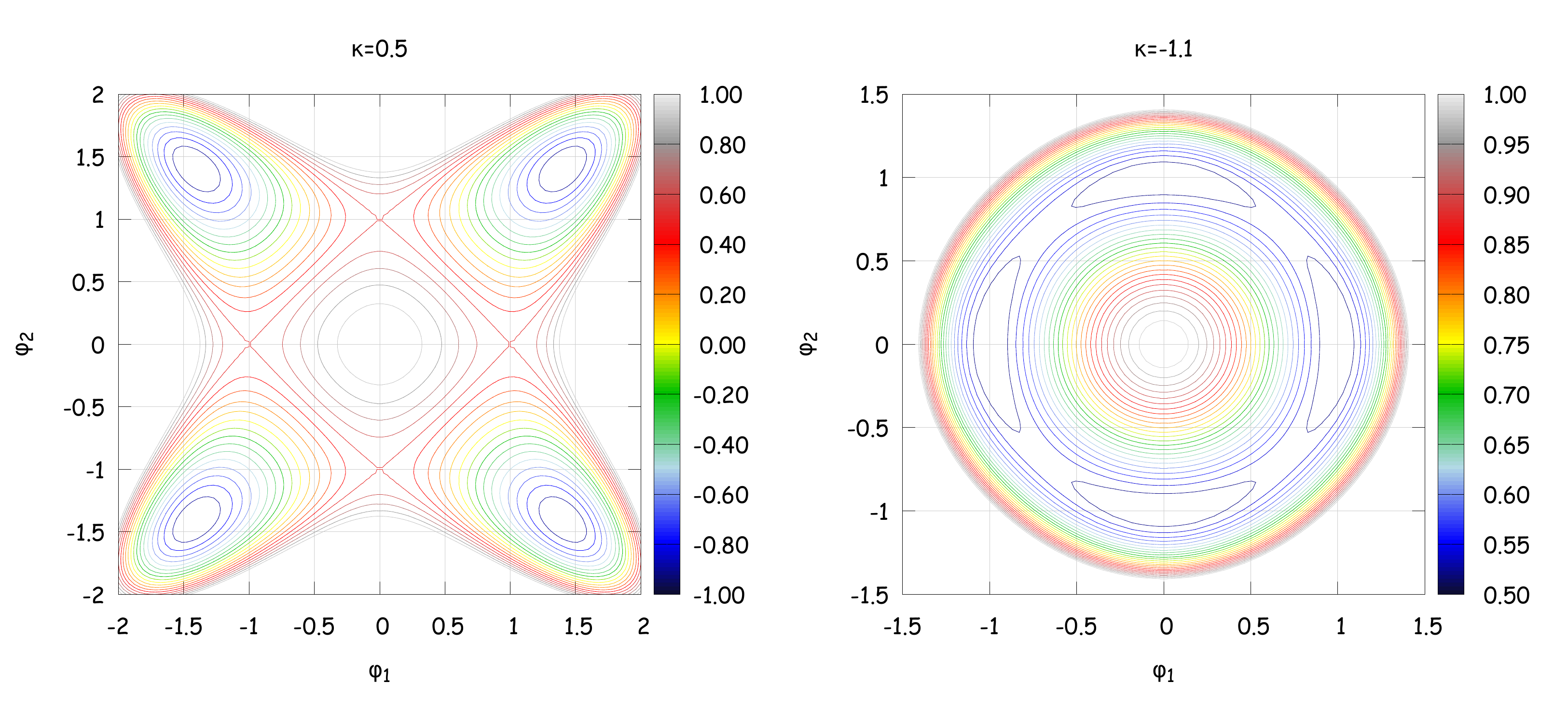}
%\end{center}
\vspace{0.5cm}
\caption{\small (Color online) Contour plot of the potential \re{pot}
for $\kappa = 0.5 < 1$ (left panel) and $\kappa = -1.1 < -1$ (right panel).
}\label{fig:1}
\end{figure}

Thus, there are kink soliton solutions in this model which interpolate between one of the vacua at $x=-\infty$
and another one at $x =+\infty$. To find these configurations explicitly we have to solve
the Euler-Lagrange equations of motion, which are the following coupled, nonlinear, partial
differential equations:
\begin{equation} \label{eq:static}
\begin{split}
     \partial_{tt}\phi_1 - \partial_{xx} \phi_1 + 2\phi_1(\phi_1^2-1-\kappa\phi_2^2) &= 0,\\
     \partial_{tt}\phi_2 - \partial_{xx} \phi_2 + 2\phi_2(\phi_2^2-1-\kappa\phi_1^2) &= 0.
\end{split}
\end{equation}

Let us consider $|\kappa| < 1$. Then a particular static kink solutions interpolate between two of four vacua:
\mbox{$(\pm 1/\sqrt{1-\kappa}, \pm 1/\sqrt{1-\kappa})$}.
So there are $2\tbinom{4}{2}=12$ types of the (anti)kinks in the model, both topological and non-topological.
However the set of discrete symmetries of the Lagrangian \re{Lag}:
\begin{subequations} \label{symm}
   \begin{align}
      x&\to-x,\\
      (\phi_1,\phi_2)&\to(-\phi_1,\phi_2),\\
      (\phi_1,\phi_2)&\to(\phi_1,-\phi_2),\\
      (\phi_1,\phi_2)&\to(\phi_2,\phi_1).
 \end{align}\end{subequations}
actually reduces the number of significantly different solitons to two.

Indeed, the solitons can be classified according to their topological charge
\be
Q_i =  \frac{1}{2} \sqrt{1-\kappa} \int \limits_{-\infty}^{\infty}
\! dx \, \frac{\partial \phi_i} {\partial x} , \qquad i=1,2.
\ee
For each component it can have three values $\{-1,0,1\}$,  kink soliton has topological charge $+1$ while
antikink has topological charge $-1$. Thus, the configuration can be labeled by the
topological charges of the components as $(n_1, n_2)$ where
$n_i\in\{-1,0,1\}$.

Let us consider the most symmetric case $\phi_2= \pm \phi_1=\phi$. Then the system of coupled equations \re{eq:static}
becomes reduced to equation
\begin{equation}
   \partial_{tt}\phi-\partial_{xx}\phi+2\phi((1-\kappa)\phi^2-1)=0.
\end{equation}
Evidently, rescaling of the field $\phi\to\tilde\phi=\phi \, \sqrt{1-\kappa}$ yields the familiar equation
of motion of the usual one-component $\phi^4$ theory. So the model \re{Lag} in this case is reduced to the
two \textit{separable} copies of the $\phi^4$ model. Static solutions ($\partial_t \phi=0$) of the system
\re{eq:static} can therefore be written as
\begin{equation}
\begin{bmatrix}
\phi_1\\\phi_2
\end{bmatrix}=\pm
\frac{\tanh\,x}{\sqrt{1-\kappa}}\begin{bmatrix}
1\\\pm1
\end{bmatrix}.
\end{equation}

The static kink solution interpolates between the vacua $-1/\sqrt{1-\kappa}$ and $1/\sqrt{1-\kappa}$
as $x$ increases from $-\infty$ to $\infty$. Similarly, the antikink solution interpolates between the stable
vacua $1/\sqrt{1-\kappa}$ and $-1/\sqrt{1-\kappa}$ as $x$ varies from $-\infty$ to $\infty$.
The corresponding 2-component configuration possess non-zero topological charge in both sectors, so there
are four kinks which correspond
to the configurations with topological charges $(1,1), (-1,1), (1, -1)$ and $(-1,-1)$, respectively. We shall refer
to this solution as \textit{double kink}.

The perturbative sector of the model consists of small linear
perturbations around one of the solutions, thus by analogy with the
usual one-component $\phi^4$ model, one can expect the process of collision between the solitons is
related to reversible exchange of energy
between the translational and vibrational modes of the individual kinks
\cite{Peyrard:1984qn,Campbell:1983xu,Anninos:1991un,Googman:2005}. However in our case there is another
possibility related to energy exchange between the sectors, so the dynamics of the kinks and the spectral
structure of the perturbations can be rather interesting.

Besides the double kink configuration,
there is another solution to the model \re{Lag} which corresponds to the topological soliton in one of
the sectors and  non-topological soliton (lump) in another sector. We shall refer to this solution as
\textit{lump kinks}. In such two-component system the
field of the topological soliton in one of the sectors interpolates between
the vacua $\mp 1/\sqrt{1-\kappa}$ and $\pm 1/\sqrt{1-\kappa}$ as before, while the non-topological soliton
belongs to the sector with topological charge 0, each of these lump solitons can be settled in one of two vacuum states,
so there are 8 such configurations labeled by the
topological charges $(1,0), (-1,0), (0, 1)$ and $(0,-1)$. For example, the static lump kink solution having
first component interpolating between the vacua $-1/\sqrt{1-\kappa}$ and $1/\sqrt{1-\kappa}$ and the second component
being the non-topological soliton about one of the vacua $\pm1/\sqrt{1-\kappa}$ is presented in
Fig.~\ref{fig:2}.

\begin{figure}
\centering
%\begin{center}
%   \includegraphics[width=12cm,angle=0]{staticSolutions_kappa=05.pdf}
   \includegraphics[width=12cm,angle=0]{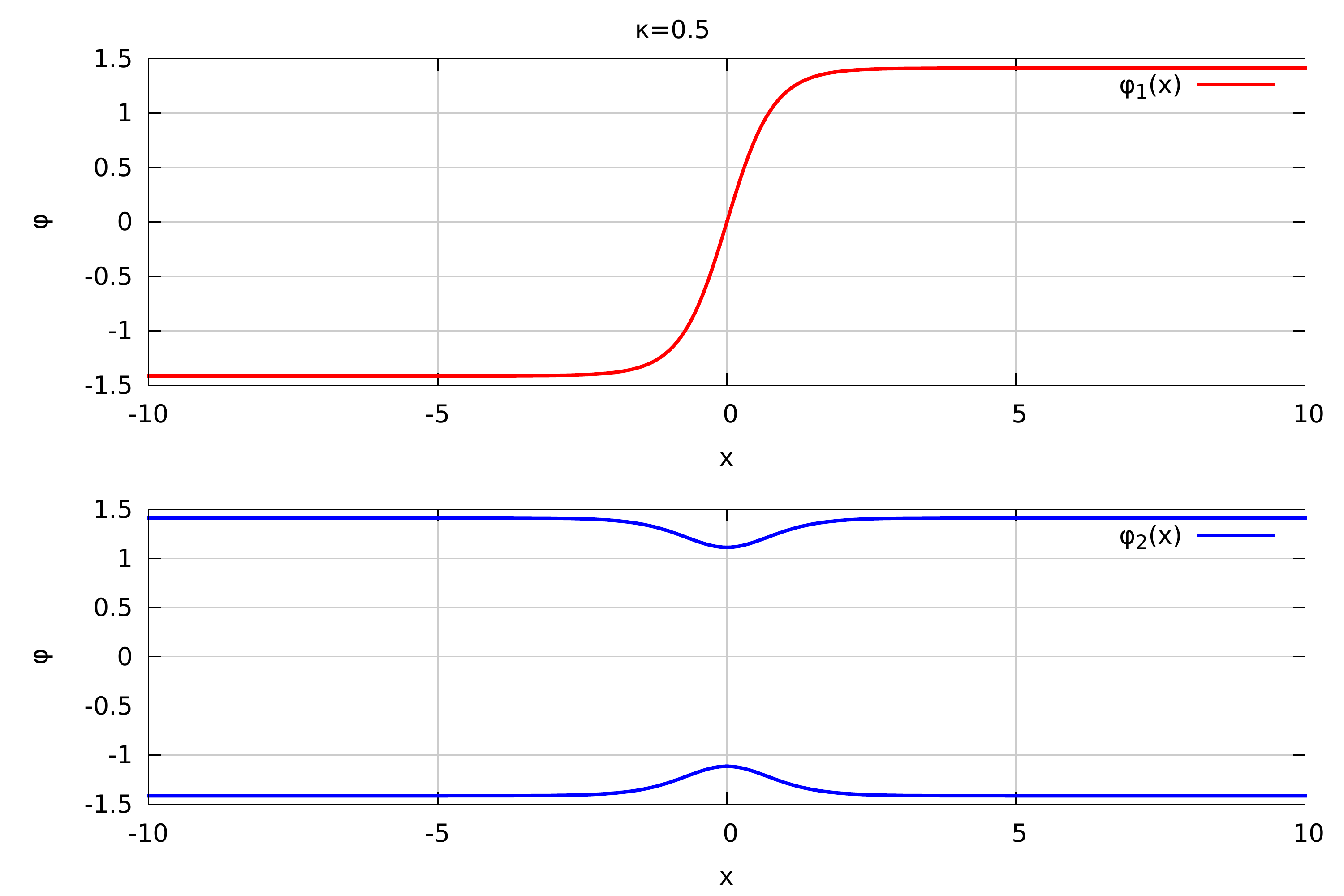}
%\end{center}
\caption{\small Lump kink static configurations $(1,0)$ is plotted for $\kappa=0.5$.
}\label{fig:2}
\end{figure}

Other static lump (anti)kinks can be obtained from this solution by using the discrete symmetries of the model
\re{symm}. Qualitatively, one can understand the physical reason of the lump kink solitons existence, if we consider the
kink configuration in one of the sectors, in the configuration space $(\phi_1,\phi_2)$ it tends to evolve
from one vacuum to another through the saddle point $(\phi_1(0)=0,\phi_2(0)=1)$. Thus, non-topological soliton in the
second sector appears. However the resulting
configuration does not reach this point exactly
because of the field derivatives contributions in the total energy. Note that although the general
structure of the solution is similar to the configuration considered in \cite{Rajaraman:1978kd}, there is no analytical
solution of the model \re{Lag}, the kink component there is rather different from the solution of the usual $\phi^4$ model.
In the case under consideration the topological solitons can be treated as "stretched out" $\phi^4$ kinks which are coupled
to the lumps in the second sector. These configurations
can be constructed numerically as solutions of the  of static equations (\ref{eq:static})
when we impose the boundary conditions
\be
\phi_1(0) = 0 ,\;\;\phi_1(\infty)=1/\sqrt{1-\kappa},\qquad
   \phi'_2(0) = 0 ,\;\;\phi_2(\infty)=1/\sqrt{1-\kappa}.
\ee

Since the coupling constant $|\kappa|<1$, we can construct an analytical approximation
of the lump kink solution considering the coupling between the components perturbatively.
The initial decoupled static
configuration consist of a kink in the first sector of the model \re{Lag} and the trivial vacuum solution
in the second sector:
\be \label{ini}
\phi^{(0)}_1 = \tanh(x); \qquad \phi_2^{(0)} = 1.
\ee
Corrections to this configuration can be obtained from the series expansion of the fields $\phi_1, \phi_2$
in powers of perturbation constant \cite{SH,Radiative}
\be  \label{pert}
\phi_1 = \phi^{(0)}_1 + \kappa \phi^{(1)}_1 + \kappa^2 \phi^{(2)}_1 + \dots;
\qquad \phi_2 = \phi^{(0)}_2 + \kappa \phi^{(1)}_2 + \kappa^2 \phi^{(2)}_2 + \dots.
\ee
The first-order time-independent corrections to the initial solution \re{ini} then can be obtained
from the equations
\be \label{firstcor}
   \begin{split}
      -\left( {\phi_1^{(1)}}\right)_{xx}+\left(6\tanh^2 x-2\right){\phi^{(1)}_1}-2\tanh x&=0,\\
      -\left( {\phi_2^{(1)}}\right)_{xx}+4{\phi^{(1)}_2}-2\tanh^2 x&=0.
   \end{split}
\ee
Evidently, as $x\to\infty$ the first order correction  $\phi_i^{(1)} \to \frac 12$
supposing that on the spacial asymptotic
 $\left({\phi_i^{(1)}}\right)_{xx} \to 0$.
This is in agreement with the vacuum shift due to perturbative coupling
in both sectors $1/\sqrt{1-\kappa}\approx1+\frac 12\kappa$.

Further, we can try to construct an analytical
approximation to the first order correction using an expansion in hyperbolic functions
\be
    \phi_1^{(1)} = \sum_{n=0}^\infty A_{n}\tanh^{2n+1}x , \qquad    \phi_2^{(1)} = \sum_{n=0}^\infty
    \frac{B_{n}}{\cosh^{2n}x}.
\ee
Substitution of these series into the first order equations \re{firstcor} yields
\begin{equation}
\begin{split}
   \phi_1^{(1)}=&\frac{16}{15}\tanh x -\frac 13 \tanh^3x-\frac{4}{75}\tanh^5x-\frac{1}{75}\tanh^7x+\cdots,\\
   \phi_2^{(1)}=&\frac 12 -\frac{1}{6\cosh^2x}-\frac{1}{28\cosh^4x}-\frac{5}{728\cosh^6x}-\frac{45}{11284\cosh^8x}+\cdots
\end{split}
\end{equation}
Similar approximation can be applied to the higher order corrections.\\

Note that the perturbative expansion reveals instability of the double kink configuration
in the case $\kappa<0$,  the system may decay into a pair of
\textit{lump kinks}. Indeed,
the Figure \ref{fig:mass} shows the masses of kinks as a function of coupling constant $\kappa$.
Evidently, for negative values of the coupling $\kappa$
the mass of a double kink is less than the mass of two lump kinks. That means that the double kink is
less energetically favorable state than two separated lump kinks.
Kinks in one-component $\phi^4$ model are stable, therefore the perturbation which can destroy the double
kink configuration must break the symmetry $\phi_1=\pm\phi_2$.
For positive $\kappa$ the double kink is energetically stable.
Moreover it is possible that the double kink could be a final state in collisions of two lump kinks
with kinks in different sectors, the outcome evidently depends on the parameters of the model, the impact velocity
and the coupling $\kappa$.  Below we discuss various solitons collisions scenarios and classify them.

\begin{figure}
\centering
%\begin{center}
   \includegraphics[width=12cm,angle=0]{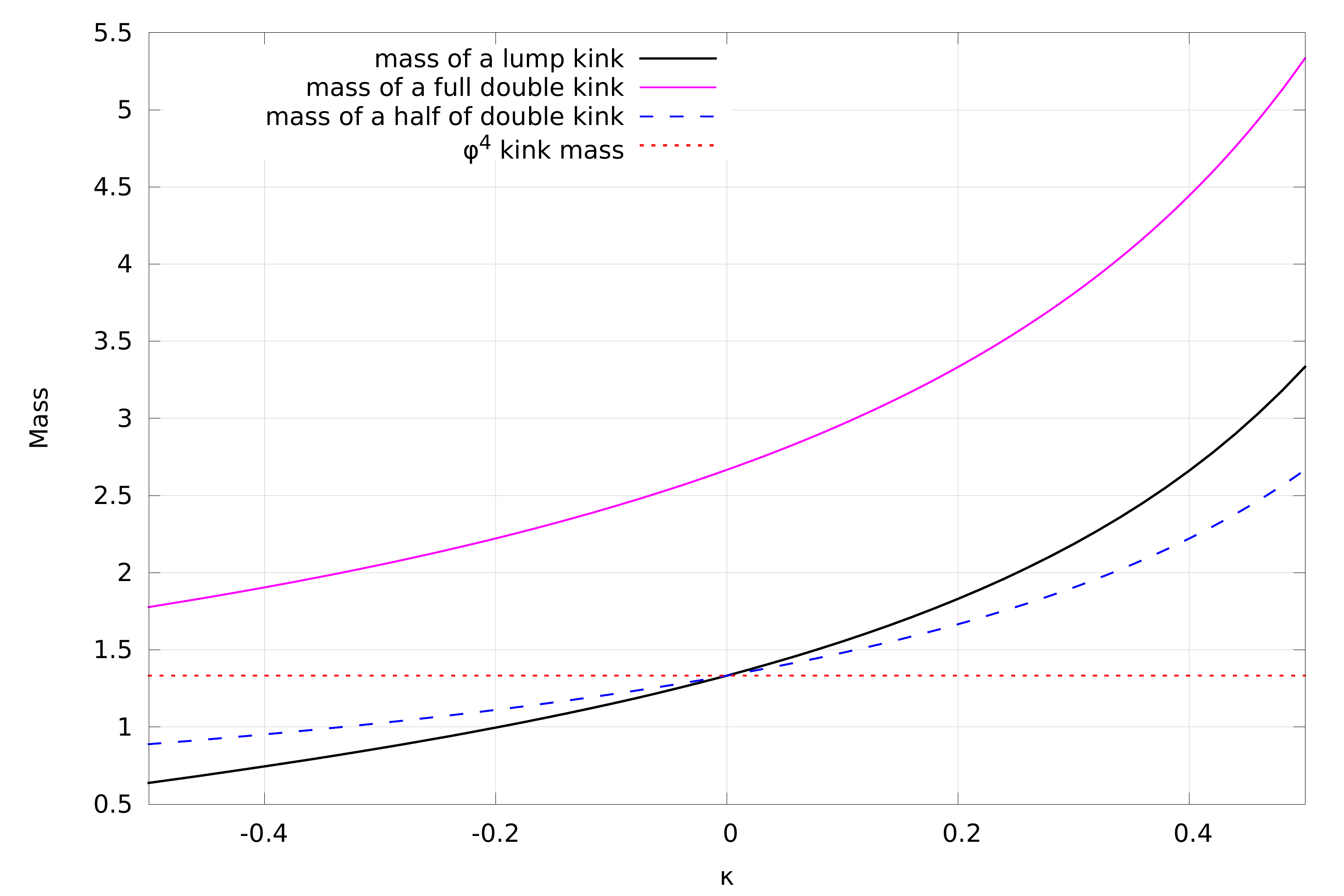}
%\end{center}
\caption{\small The mass of the two kink types, the lump kink and the double kink,
as function of coupling constant $\kappa$.
}\label{fig:mass}
\end{figure}

\section{Kink-antikink collisions in two-component $\phi^4$ model}

A phenomenon known as the two-bounce resonance soliton scattering in the usual one-dimensional $\phi^4$ model
was observed by many researchers in numerical simulations
\cite{Peyrard:1984qn,Campbell:1983xu,Anninos:1991un,Googman:2005}.
Unfortunately there is no complete analytic description for such  process, the model is non-integrable
and the dynamics of the solitons is highly complicated.
Numerical simulations performed over last 20 years
\cite{Peyrard:1984qn,Campbell:1983xu,Anninos:1991un,Googman:2005}
discovered an interesting result: the process of collision between the solitons is of fractal nature,
depending on the impact velocity, it may produce various results and
for some range of values of the impact velocity the $K\bar K$ pair form a set of bouncing windows.

Although  semiclassical consideration
\cite{Campbell:1983xu,Anninos:1991un,Googman:2005} could capture the most important
features of the process, the analysis entirely rely on extensive numerical computations.
It is known that for initial velocities of a kink and an antikink above a critical value
$v_{cr}=0.2598$
two incident solitons always escape to infinity after
collision, with the emission of some radiation, thus the process is similar to quasielastic scattering of two particle-like
objects \cite{Peyrard:1984qn,Campbell:1983xu,Anninos:1991un,Googman:2005}.  For the impact velocities below
$v_{cr}$, the kink-antikink pair generically becomes trapped, it annihilates into radiation, however numerical analysis
reveals the fractal structure with a sequence of narrow bouncing windows
within which the solitons are again able to escape to infinity.
The first of two-bounce window opens at the impact velocity $v_{in}=0.189$ then the
resonance windows become more narrow as the impact velocity increases.

If we look on the edge of a two-bounce window, we can observe a narrow sequence of the 3-bounce windows which
are characterized by the same relations between the windows widths and the impact velocities (self-similar structure)
\cite{Anninos:1991un}. These windows are called 3-bounce ones.
Furthermore, on the edge of every 3-bounce window we can find the 4-bounce windows structure, etc.
However, the fractal dimension of this dynamical structure of the collision of the solitons
is restricted, so we refer to this to as a quasi-fractal structure.

The accepted explanation of the appearance of these windows is that they are related to reversible exchange of energy
between the translational and vibrational modes of the individual kinks. At the initial impact, some kinetic
energy is transferred into internal shape modes of the kink and antikink. They then separate and propagate
almost independently, but for initial velocities less than $v_{cr}$ they no longer have enough translational energy to
escape their mutual attraction, and so they return and collide one more time \cite{Campbell:1983xu,Anninos:1991un}.
At this point some of the energy
stored in the shape modes can be returned to the translational modes,
tipping the energy balance back again and allowing the kink and antikink to escape to infinity,
provided that there is an appropriate resonance between the time interval between the two collisions, and the
period of the internal modes. More generally, sufficient energy might be returned to the translational modes
after three or more kink-antikink collisions, leading to an intricate nested structure of resonance windows.
Thus, the resonance condition is that the time $T$ between two consequent collisions should satisfy the equation
\cite{Campbell:1983xu,Anninos:1991un}
\begin{equation} \label{time2coll}
\omega_s T=\delta + 2 \pi n
\end{equation}
where $\delta$ is a phase, and $n$ is an integer \footnote{Note that it was shown recently
\cite{Dorey:2011yw} that
the resonance windows in a model with a polynomial potential may appear
due to existence of the collective bound states trapped by the kink-antikink pair.
}.
It can be shown that $\frac{2\pi}{\omega_s}=5.13$ and $\delta=3.3 (\approx \pi)$. The relation (\ref{time2coll}) can be modulated \cite{HS}.

This mechanism allows us to construct an  effective model of the collision process \cite{Campbell:1983xu}.
The idea is to separate collective coordinates of the kink-antikink system and make use
of the Yukawa interaction between the well separated pair mediated by the meson state of mass $m$.
Surprisingly enough, this model works  rather well.
\begin{figure}
\centering
\includegraphics[width=0.75\linewidth,angle=0]{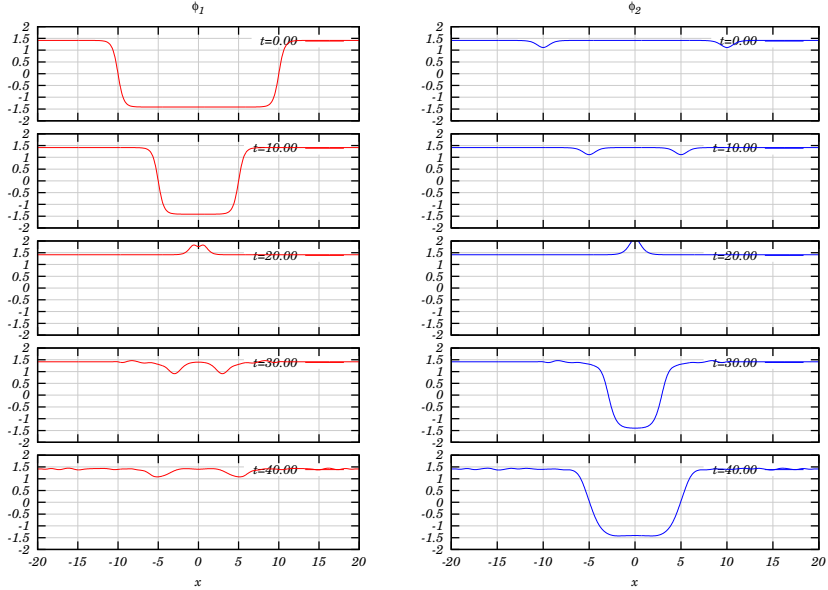}
\caption{\small Collision of two lump kinks
$
(-1,0)+(1,0)\to(0,-1)+(0,1)
$
at $\kappa=0.5$ and initial velocity
$v_{in}=0.5$.}\label{fig:3}
\end{figure}

Our goal is to analyse the process of collision of the soliton configurations in two-component model
\re{Lag} numerically.
The initial data used in our simulations represent widely separated kinks or lump kinks propagating from both
sides towards a collision point, the parameters of the collision process are the impact velocity $v$ and the coupling
constant $\kappa$.
To find a numerical solution of the PDE describing the evolution of the system, we used the
pseudo-spectral method on a discrete grid containing 2048 nodes with periodic boundary conditions.
For the time stepping function we used symplectic (or geometric) integrator of
8th order to ensure that the energy is conserved.
The time and the spacial steps are $\delta x = 0.25$ and $\delta t = 0.025$, respectively, so
the numerical errors are  of order of $(\delta t)^8$.

Evidently, if the sectors of the two-component model \re{Lag} are decoupled, the pattern of collision reproduces
the chaotic structure we described above.

\subsection{Lump kink collisions - same sectors}
Let us consider collision of two lump kinks $(-1,0)+(1,0)$ at $\kappa = 0.5$. The numerical simulations
reveal rather interesting pattern,
for initial velocities of a lump kink and an lump antikink above a critical value
$v_{cr}=0.453$, an intermediate state which is
created in both sectors at the collision center, decays into lumps in the first sector and into kinks in the second one.
Then two solitons escape to infinity, with the emission of some radiation, so the process is similar to
quasielastic scattering of two $\phi^4$ kinks up to the flip between sectors:
$$
(-1,0)+(1,0)\to(0,-1)+(0,1).
$$
This process is displayed at Fig.~\ref{fig:3}. For the impact velocities below
$v_{cr}$ we the observe the resonance energy exchange between the sectors
with a sequences of bouncing windows within which the solitons are again able to escape to infinity,
as seen in Fig.~\ref{fig:4}.
\begin{figure}
\centering

\includegraphics[width=0.85\linewidth,angle=0]{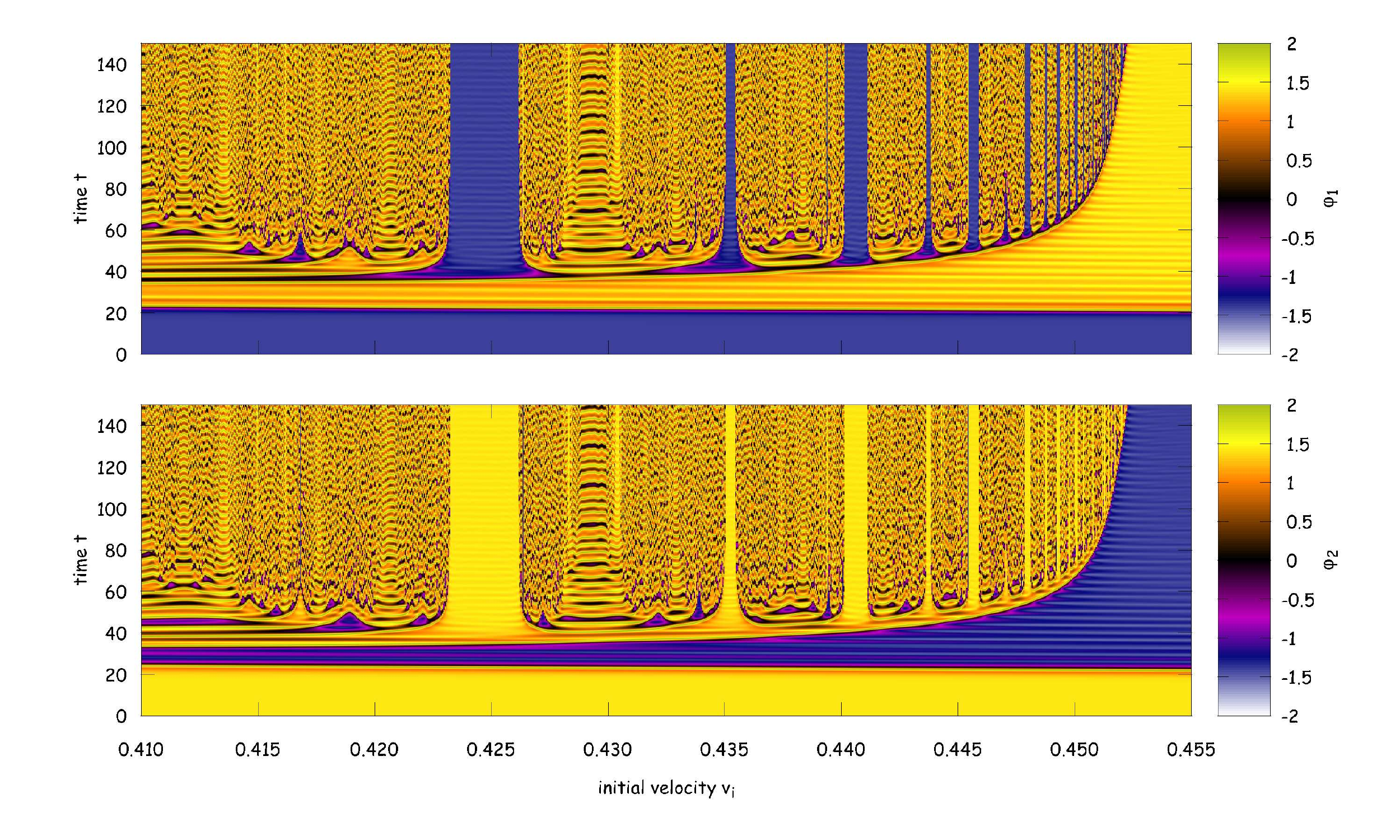}%{collisions_kappa=05.pdf}
\vspace{0.2cm}
\caption{\small (Color online) Two lump kinks $(-1,0)+(1,0)$ collision at $\kappa=0.5$.
The plots represent the field values measured at the collision center
as function of impact velocity and time. Upper and bottom plots
represent the $\phi_1$ and $\phi_2$ components, respectively.
}\label{fig:4}
\end{figure}
The first of two-bounce windows open at the impact velocity $v_{in}=0.423$, the
consequent structure of the resonance windows is not so regular as in the case of the usual $\phi^4$ model, also
some false windows appear. Evidently, this can be related to more complicated structure of the spectrum of
excitations, alongside with the internal mode of the kinks there are degrees of freedom associated with excitations
of the lumps, also some collective degrees of freedom in the system kink-lump may appear.

In all windows we observed the same permutation of the kink and the lump sectors in the final state.
For smaller impact velocities
and between the windows, the intermediate state which is created in both sectors at the collision center,
decays into two oscillons (one in each sector) with some amount of radiation emitted.
The final states of the process then are:
\begin{equation}
(1,0)+(-1,0)\to
\begin{cases}
   (1,0)+(-1,0)\;\;&\text{bounce~~~~(I)}\\
   (0,1)+(0,-1)&\text{flip~~~~(II)}\\
   (0,0)&\text{annihilation into oscillon~~~~(III)}\,.
\end{cases}
\end{equation}
However for ultra-relativistic impact velocities (for example at $v=0.950,\,\kappa=0.386$, see Fig.~\ref{fig:SameZeros}),
we also found another final state
\begin{equation}
 (1,0)+(-1,0)\to(-1,-1)+(1,1)~~~~\text{(IV)}.
\end{equation}
In this process relativistic collision of the lumps leads to the production of the kinks in the second sector.
They bind to the kinks from the first sector and then the pairs travel as double kinks.

In Fig.~\ref{fig:SamePhaseDiag} we present the final vacuum state in the center of collision 
after a long time $\phi_i(x=0, t=150)$ as a function of the parameters of the process of collision $(1,0)+(-1,0)$, 
the value of the coupling constant $\kappa$  and the initial velocity $v_i$. 
Before the collision the initial value of the field at the center was taken as $\phi_1(0,0)=\frac{1}{\sqrt{1-\kappa}}$ 
(the only possiblity for such configuration) and $\phi_2(0,0)=-\frac{1}{\sqrt{1-\kappa}}$ (one of two possibilities). 
We will denote the state of that vacuum in square brackets as $\Phi=\sqrt{1-\kappa}[\phi_1,\phi_2]=[1,-1]$.
The final state of the vacuum in the collision center then can be defined as $\Phi=[-1,-1]+2Q$, 
where $Q$ is the topological charge of the kink moving towards asymptotic $x\to -\infty$.

Depending on the parameters of the collision process, we observed various scenario. 
If the final vacuum state remains the same after collision,  the kinks are bounced back 
(cf. Fig.~\ref{fig:SameZeros}, panel (I), where we plotted positions of the topological zeros for that collision). 
On the upper left plot in Fig.~\ref{fig:SamePhaseDiag} the bouncing corresponds to the regions colored in blue. 
The largest of these regions has a shape of a triangle at  
$\kappa=0$, note that the bouncing is also observed for some set 
of higher impact velocities and relatively strong values of the 
coupling $\kappa$, both positive and negative.

Black regions in Fig.~\ref{fig:SamePhaseDiag}  correspond to the final state $[-1,1]$, in other words 
the kink propagating on the left towards the spacial asymptotic $x \to -\infty$ after the collision 
now belongs to the sector with 
topological charge  $Q=\frac{1}{2}([-1,1]-[-1,-1])=(0,1)$. Thus, the collision leads to the topological fliping 
between the sectors (cf. Fig.~\ref{fig:SameZeros}, panel (II)).

The regions  in Fig.~\ref{fig:SamePhaseDiag} which are colored in red, correspond to final state of the configuration 
in the region between the collision center and the asymptotic $x \to -\infty$ with total topological charge equal
zero. This is the process of $K \bar K$ annihilation which dominates at relatively small initial velocities
(cf. Fig.~\ref{fig:SameZeros}, panel (III)).

Finally, we observed that for the relativistic impact velocities and relatively large  positive values 
of the coupling constant $\kappa$ the process of the production of the double kink is allowed  
as illustrated in Fig.~\ref{fig:SameZeros}, panel (IV). The corresponding regions 
belong to the sector with topological charge  $ Q=(1,1)$, 
 in Fig.~\ref{fig:SamePhaseDiag} they  are colored in yellow. 

Note that the  borders between those regions are very complicated, presumably they are demonstrating some kind of 
fractal structure.  

\begin{figure}
\centering
\includegraphics[width=0.8\linewidth,angle=0]{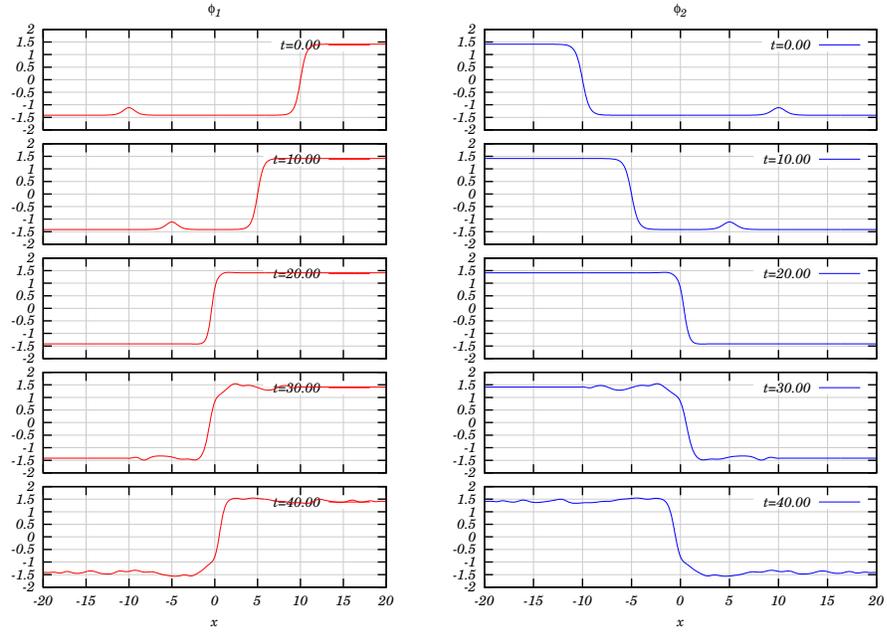}
\vspace{0.2cm}
\caption{\small (Color online) Collision of two lump kinks  in different sectors mediated
by the lumps at $\kappa=0.5$.}\label{fig:5}
\end{figure}

\begin{figure}
\centering
\includegraphics[width=0.8\linewidth,angle=0]{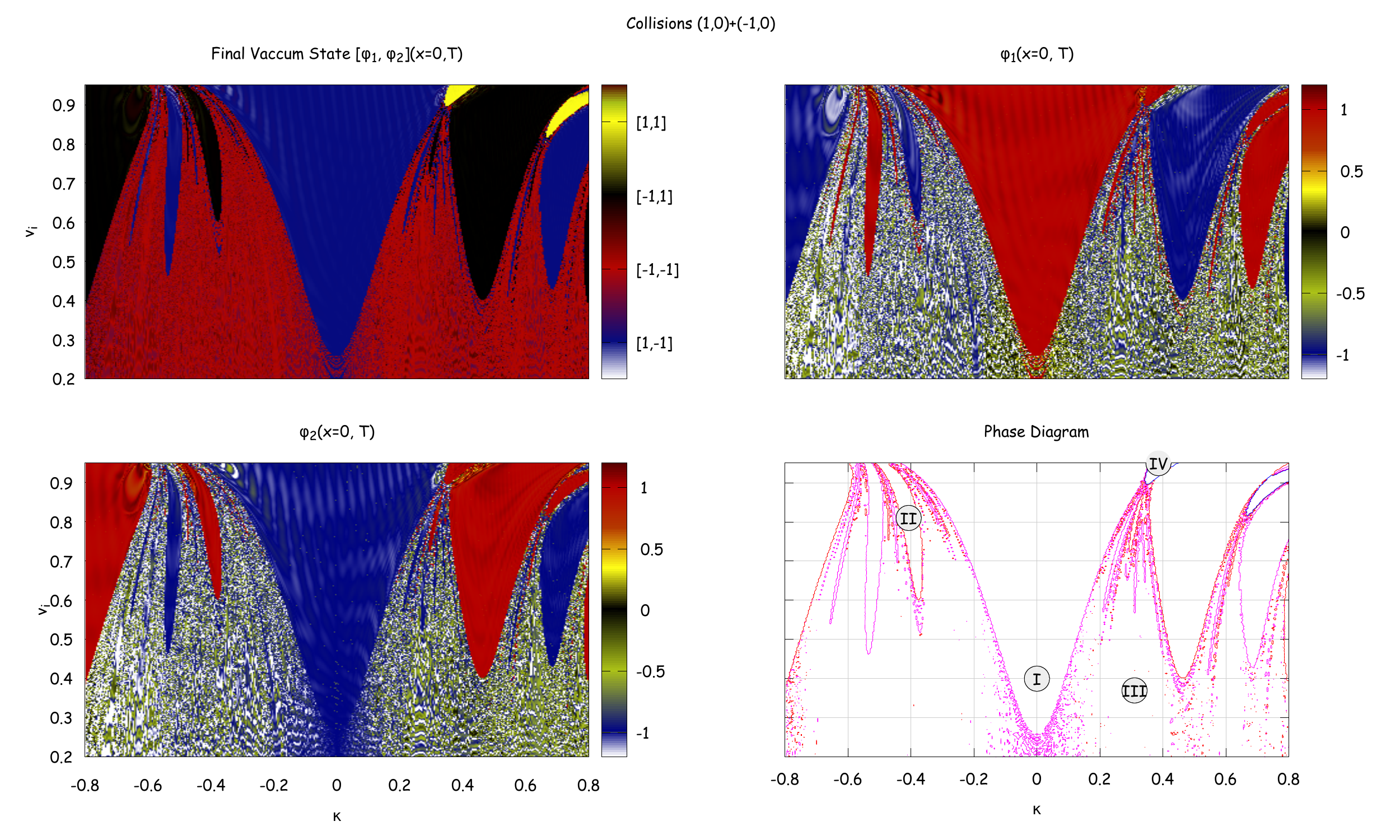}
\vspace{0.2cm}
\caption{\small (Color online) Final state of the two lump kinks collision $(1,0)+(-1,0)$.
}\label{fig:SamePhaseDiag}
\end{figure}

\begin{figure}
\centering
\includegraphics[width=0.8\linewidth,angle=0]{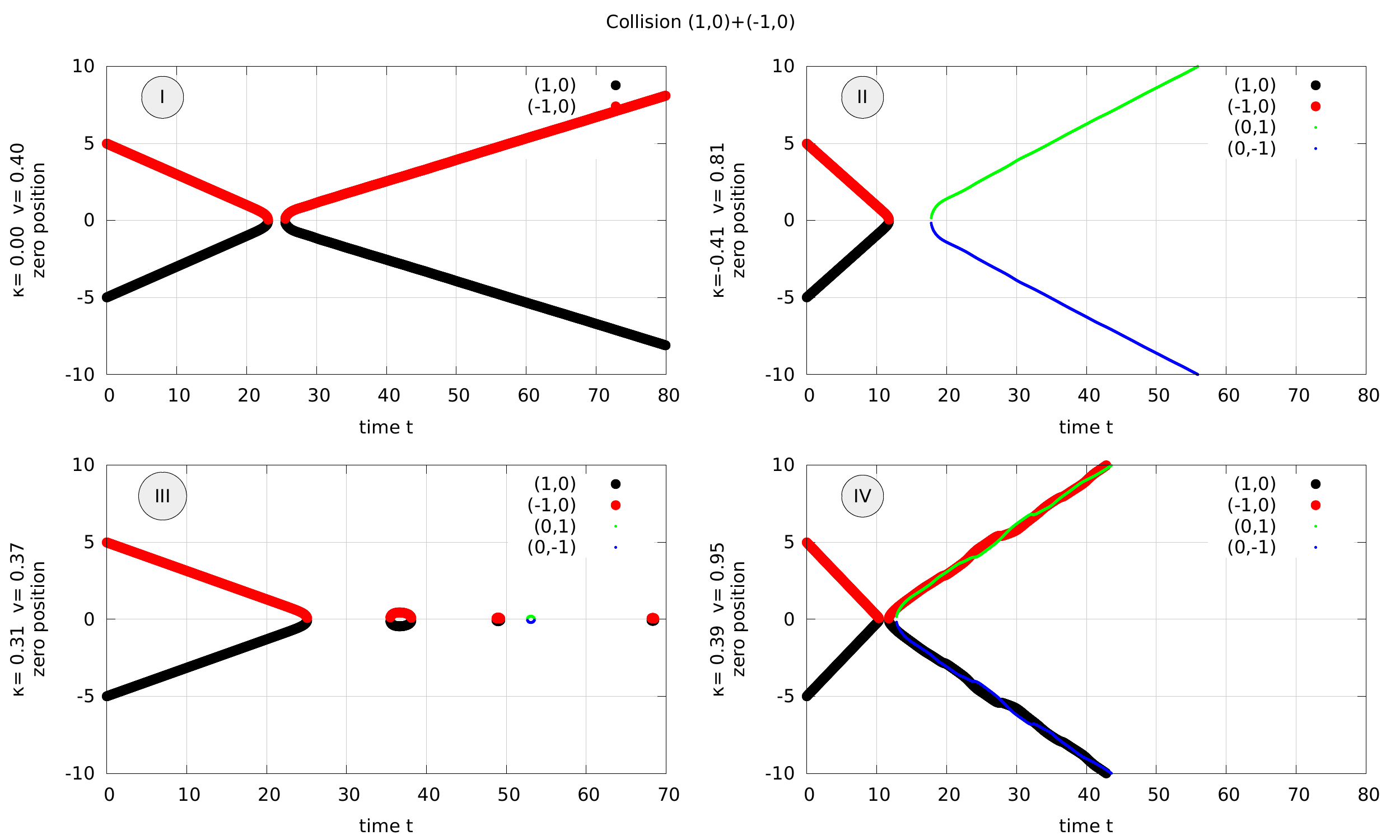}
\vspace{0.2cm}
\caption{\small (Color online) Example of topological zeros motion for two lump kinks collision $(1,0)+(-1,0)$ in the same sectors.
(I) bounce $(1,0)+(-1,0)$, (II) flip $(0,-1)+(0,1)$, (III) annihilation $(0,0)$, (IV) double kinks creation $(-1,-1)+(1,1)$.
}\label{fig:SameZeros}
\end{figure}

\subsection{Lump kink collisions - different sectors}
Since the kink is complemented by a non-topological partner in the opposite sector, there is another process
of collision between the kinks from {\it different sectors}, mediated by the lumps. Numerical simulations reveal that
at relatively small velocities $v < 0.565$ the energy of collision between the kink and the lump
is transformed into excitations of the kinks which are erratically oscillating around the collision center
it the both sectors, see Fig.~\ref{fig:5}. Thus the resulting configuration is actually an excited double kink.
It will be shown in the next section that the double kink for $\kappa>0$ has at least two oscillational modes.
One of them corresponds to the oscillational mode of the $\phi^4$ kink and has the same frequency $\sqrt{3}$.
There is also a mode which is responsible for translational oscillations of the kinks in both sectors around the center of mass.
Both of these frequencies can be seen in the power spectrum of the field measured in the center of mass.
Moreover due to nonlinearities combinations of those frequencies are also visible (cf. Fig.~\ref{fig:fft1} ).

\begin{figure}
\centering
\includegraphics[width=0.8\linewidth,angle=0]{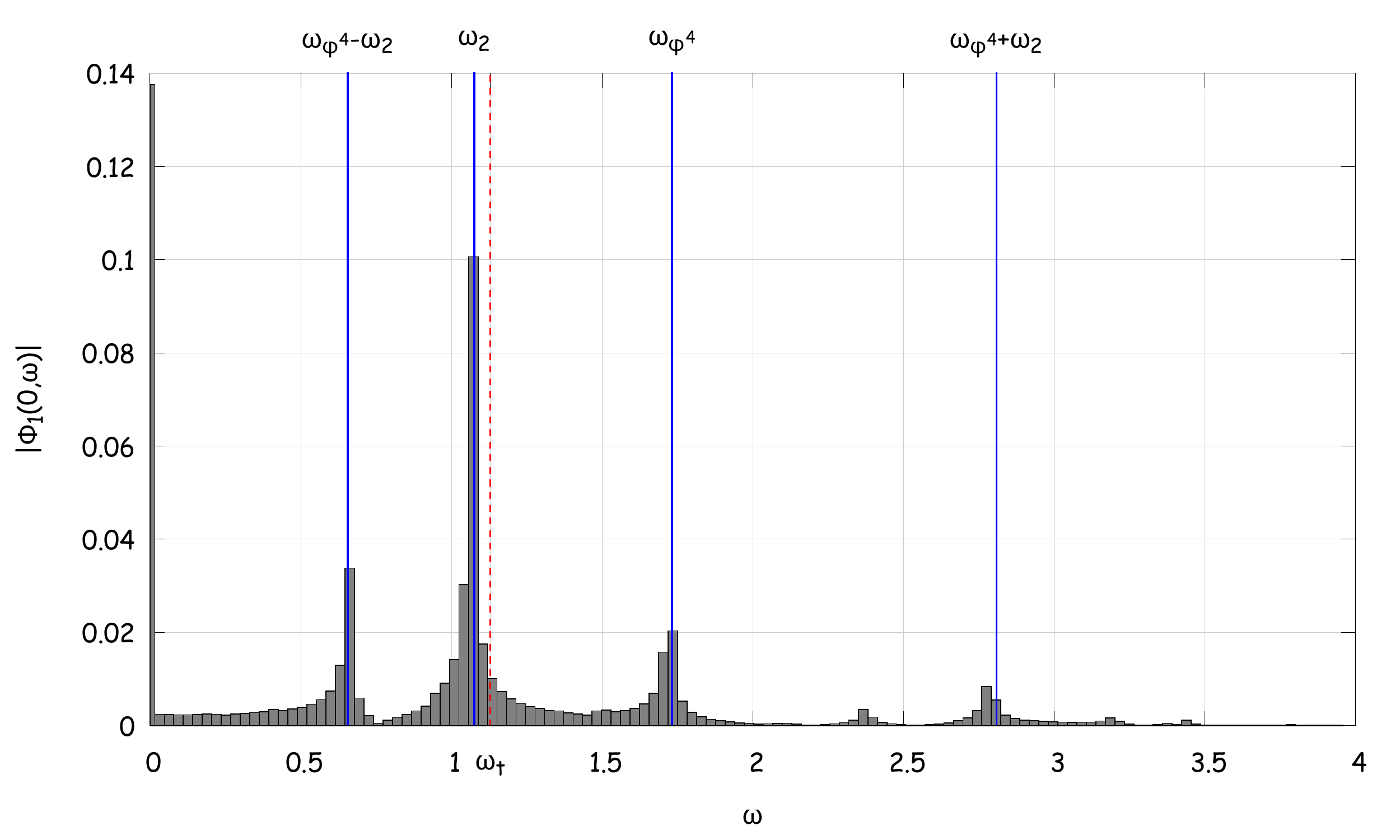}
\vspace{0.2cm}
\caption{\small (Color online) Power spectrum of the field in the first sector in the collision center for
$(1,0)+(0,-1)\rightarrow(1,-1)$ and $\kappa=0.5$ at $v_{in}=0.2$.
As the result of the collision an excited double kink is created.
One can identify the frequency of the usual $\phi^4$ internal mode $\omega_{\phi^4}=\sqrt{3}$, another 
frequency $\omega_t=\sqrt{-\frac{5}{2}+\frac{1}{2}\sqrt{\frac{25+7\kappa}{1-\kappa}}}=1.129$
(see the next section for details) is lowered because of nonlinearities to $\omega_2=1.077$, other combination of those frequencies.
}\label{fig:fft1}
\end{figure}
For impact velocity $v=0.565$ we observed the bouncing of the kinks, a narrow resonance window opens,
a sequence of the bouncing windows of decreasing width follows it up to the $v_{cr}=0.584$.
This resonance structure is much more regular than
in the case of the collision of two lump kinks in the same sector, there is no false windows and the dependence of the
windows on the impact velocity closely reminds the well known pattern observed in the usual one-component $\phi^4$ theory
\cite{Campbell:1983xu,Anninos:1991un}. Evidently we can consider it as undirect evidence of the simple mechanism of
the resonance energy
exchange between the kink and the lump which involves only the internal mode of the kink,
there is no energy exchange between sectors.

Above this threshold velocity the lump kinks
are passing through the lumps associated with the kinks from the second sector, thus this is a regime of quasi-elastic
collision, see Fig.~\ref{fig:6}.

Variations of the parameters of the collision process, the
coupling constant $\kappa$ and the impact velocity $v$ reveal several possible final states:
 \begin{equation}
(1,0)+(0,-1)\to
\begin{cases}
   (1,-1)&\text{double kink creation~~~~(I)}\\
   (0,-1)+(1,0)\;\;&\text{passage~~~~(II)}\\
   (1,0)+(0,-1)&\text{bounce~~~~(III)}
\end{cases}
\end{equation}
Note  that the double kink production is allowed only for $\kappa>0$, we already mentioned that for $\kappa<0$ the
double kinks are unstable with respect to decay into  the lump kinks.

All the processes decribed above are illustrated in Fig.~\ref{fig:DiffPhaseDiag2} which 
presents the final state of the collision $(1,0)+(0,1)$ associated to the kink on the left hand side from the 
collision center propagating towards the spacial asymptotic $x\to -\infty$ as   
$\Phi=[-1,-1]+2Q$. Similar to the Fig.~\ref{fig:SamePhaseDiag}, the 
black regions $\Phi=[-1,1]$ correspond to the quasi-elastic scattering of the kinks (passage), in this case the topological
charge of the scattered kink in the final state is 
$Q=(1,0)$ (cf. Fig.~\ref{fig:DiffPhaseDiag2},  (II)).
Then the regions in Fig.~\ref{fig:DiffPhaseDiag2} which are colored in
blue, correspond to the bouncing of the kinks, then in the final state we have $Q=(1,0)$ (cf. Fig.~\ref{fig:DiffPhaseDiag2},
regions (III) and (IV)).
Note that there is another possibility, as a result of the collision two kinks can merge 
forming an (excited) double kink. These processes are depicted as whitish, yellow and light blue colored areas  
on the right side of the plot (Fig.~\ref{fig:DiffPhaseDiag2}, (I)).

\begin{figure}
\centering
\includegraphics[width=0.8\linewidth,angle=0]{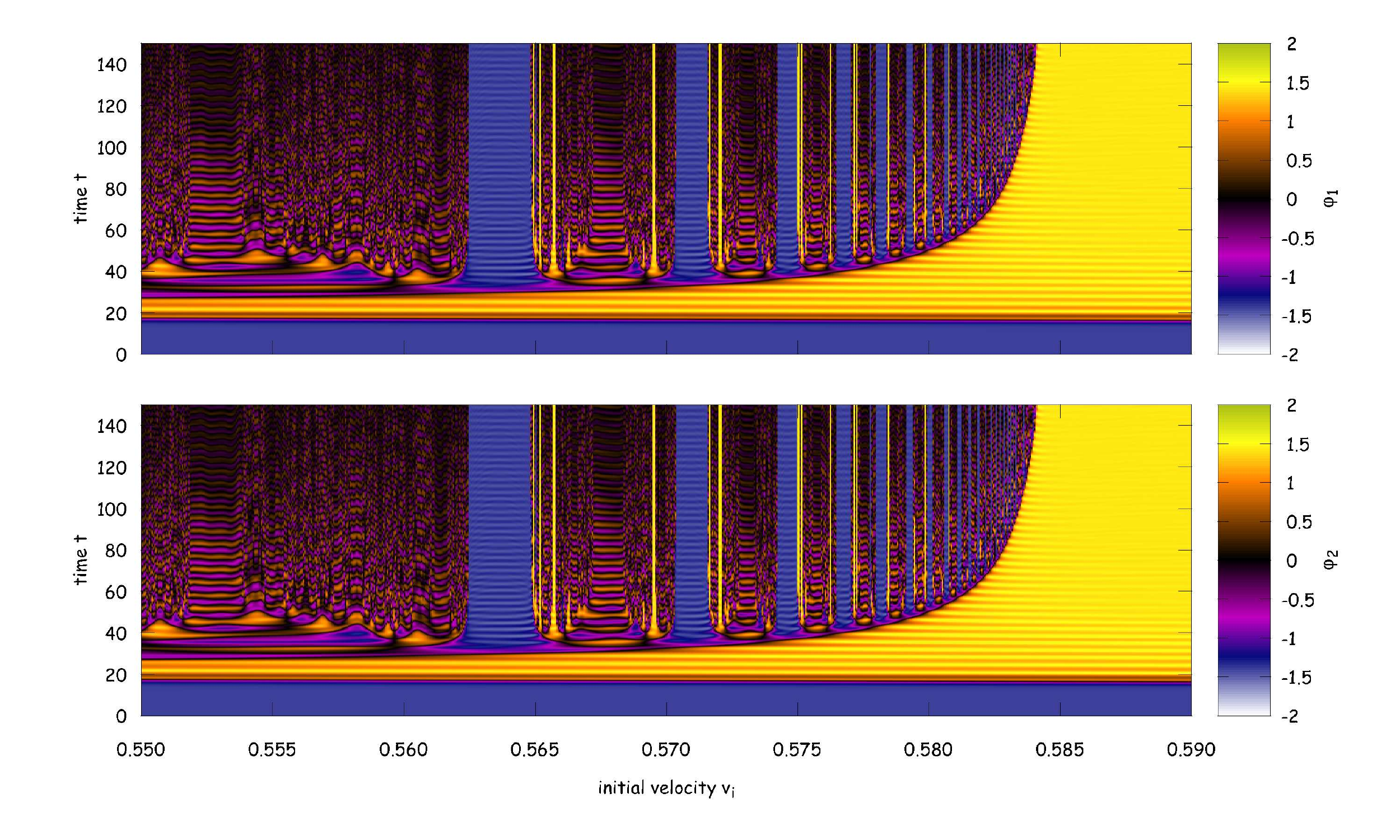}%{collisions_kappa=05_2polar.pdf}
\vspace{0.2cm}
\caption{\small (Color online) Collision of two lump kinks in different sectors mediated
by the lumps at $\kappa=0.5$.
The plots represent the field values measured at the collision center
as function of impact velocity and time. Upper and bottom plots
represent the $\phi_1$ and $\phi_2$ components, respectively.
}\label{fig:6}
\end{figure}

\begin{figure}
\centering
\includegraphics[width=0.8\linewidth,angle=0]{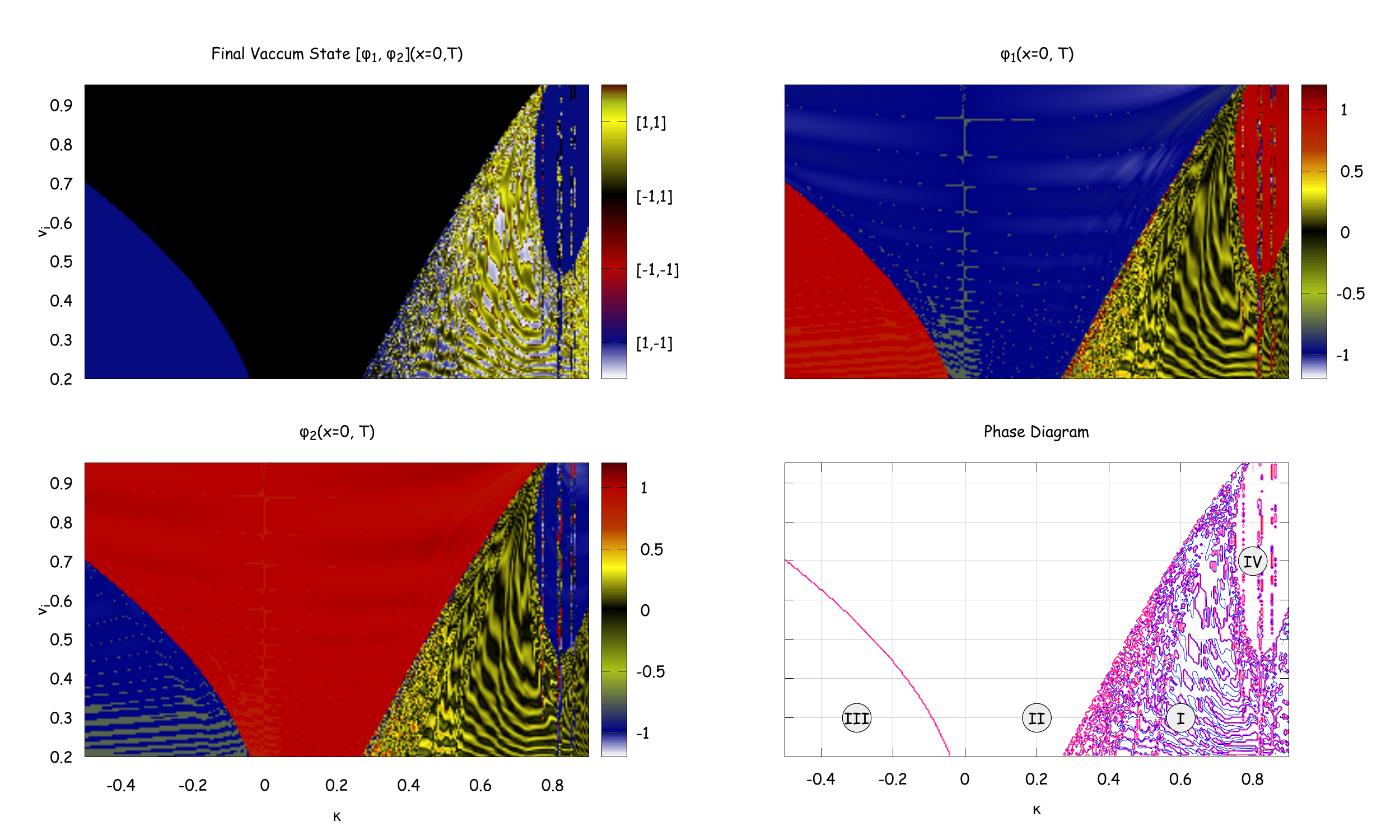}
\vspace{0.2cm}
\caption{\small (Color online) Final state of two lump kinks collision $(1,0)+(0,1)$ in different sectors.
}\label{fig:DiffPhaseDiag2}
\end{figure}

\begin{figure}
\centering
\includegraphics[width=0.8\linewidth,angle=0]{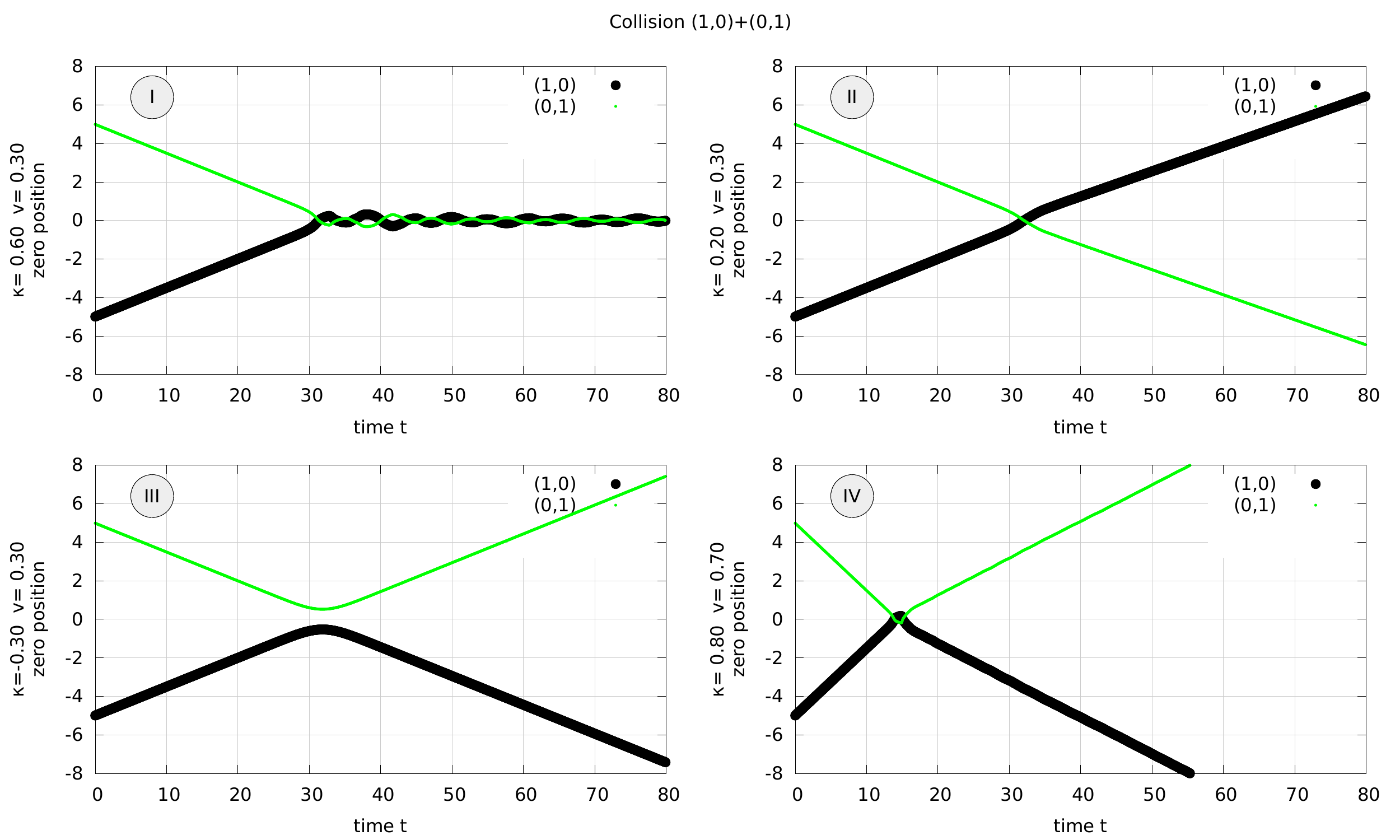}
\vspace{0.2cm}
\caption{\small (Color online) Example of topological zeros motion of two lump kinks collision $(1,0)+(0,1)$ in different sectors.
(I) double kink creation $(1,1)$, (II) passage $(0,1)+(1,0)$, (III) bounce $(1,0)+(0,1)$, (IV) multibounce  $(1,0)+(0,1)$.
}\label{fig:DiffZeros2}
\end{figure}

\subsection{Double kink - lump kink collisions}
Another interesting solitonic collision process in the two-component model \re{Lag}
is the double kink - lump kink collision. Let us take the
initial state as, for example  $(-1,-1)+(1,0)$. Since the total topological charge in the first sector
is 0 we could expect that the outcome of the collision, at least at small impact velocities, will be
just annihilation of the lump kink and one of the components of the double kink.
However, a single component of the double kink is not a solution of the model \re{Lag},
so it has to deform into the lump kink then. Indeed, at relatively small impact velocities $v<0.76$ we observed such a process
$(-1,-1)+(1,0) \to (0,-1) + (0,0)$,
besides radiation and oscillon remains in the first sector as the result of annihilation of the kink-antikink components.

It is known that the resonance excitation of the oscillon can produce $K\bar K$ pairs \cite{Romanczukiewicz:2010eg}, so
at higher energies other scenarios are possible. Indeed at $\kappa = 0.5$, within the impact velocity
range $0.76<v<0.82$ we observed another resonance structure between 3 channels
\begin{equation}
(-1,-1)+(1,0)\to
\begin{cases}
   (0,-1)+(0,0)\\
   (0,-1)+(0,1)+(0,-1)\\
   (0,-1)+(1,0)+(-1,0)
\end{cases}
\end{equation}
The results of numerical simulations are presented in Fig.~\ref{fig:7} where we plotted final velocity of
the solitons vs the impact velocity. The bottom plot represents zoom in on the smaller subrange of the impact
velocities in between,
evidently this plot is qualitatively similar to the upper plot. Thus a new type of self-similar fractal structure
appears. Note the width of the windows {\it increases} with the increasing of the impact velocity, so this structure
is inverted with respect to the usual fractal structure observed in the $\phi^4$ model \cite{Campbell:1983xu,Anninos:1991un}.
The complicated fractal structure probably is possible due to the fact that the double kink has at
least two bound modes which can couple to each other and to the oscillational mode of the lump kink.
Since at the collision process the double kink behaves as a bound state of the lump kinks,
this type of collision can be interpreted as a three body problem.

Note that for $\kappa<0$ the double kink is unstable, however the perturbation due to the lump kink is relatively weak, so
the double kink is moving as a single object and it decays into the pair of lump kinks
just before or already during the collision.

\begin{figure}
\centering
\includegraphics[width=0.8\linewidth,angle=0]{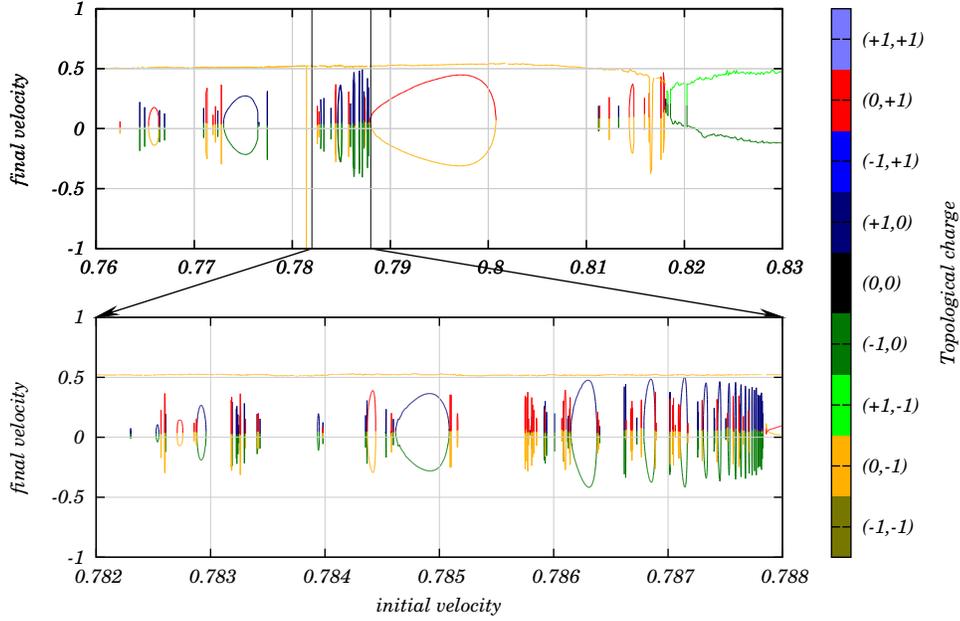}
\vspace{0.2cm}
\caption{\small (Color online) Double kink - lump kink $(-1,-1)+(1,0)$ collision  at $\kappa=0.5$.
The plots represent the escape velocity  of different solitons produced in both sectors after collision as a function of impact velocity.  Bottom plot represent the zoomed in region of the upper plot.
}\label{fig:7}
\end{figure}
Finally, as the impact velocity increases above the critical value $v_{cr}=0.82$, quasi-elastic scattering with flipping
of the sectors is observed:
\begin{equation}
   (-1,-1)+(1,0)\to(+1,-1)+(-1,0)
\end{equation}
thus, the total topological charge in each sector remains the same but the kink component of the lump kink
is getting captured into the double-kink state while the second component of the former configuration becomes
the lump kink.

In Fig.~\ref{fig:DoublePhaseDiag3} we have gathered our results for the various channels 
of the double kink - lump kink collison  $(1,1)+(-1,0)$. Here we present the values of the fields
at the center of collision, evidently the spectrum of the final states is much richer   
than in the case of the lump kink collisions we discussed above.
The colors which represent the final state in Fig.~\ref{fig:DoublePhaseDiag3}, are 
similar to the palette we used in the Figs.~\ref{fig:SamePhaseDiag},\ref{fig:DiffPhaseDiag2} above. 
Note that since 
the double kink and the lump kink has different masses, the center of collision is not 
a center of mass of the system, so the dynamics becomes rather involved. For the sake of completeness
in Fig.~\ref{fig:DoubleZeros3} we also 
presented various examples of the evolution of the topological zeros, associated with the position of the solitons 
in the process of the double kink-lump kink  collision.

\begin{figure}
\centering
\includegraphics[width=0.8\linewidth,angle=0]{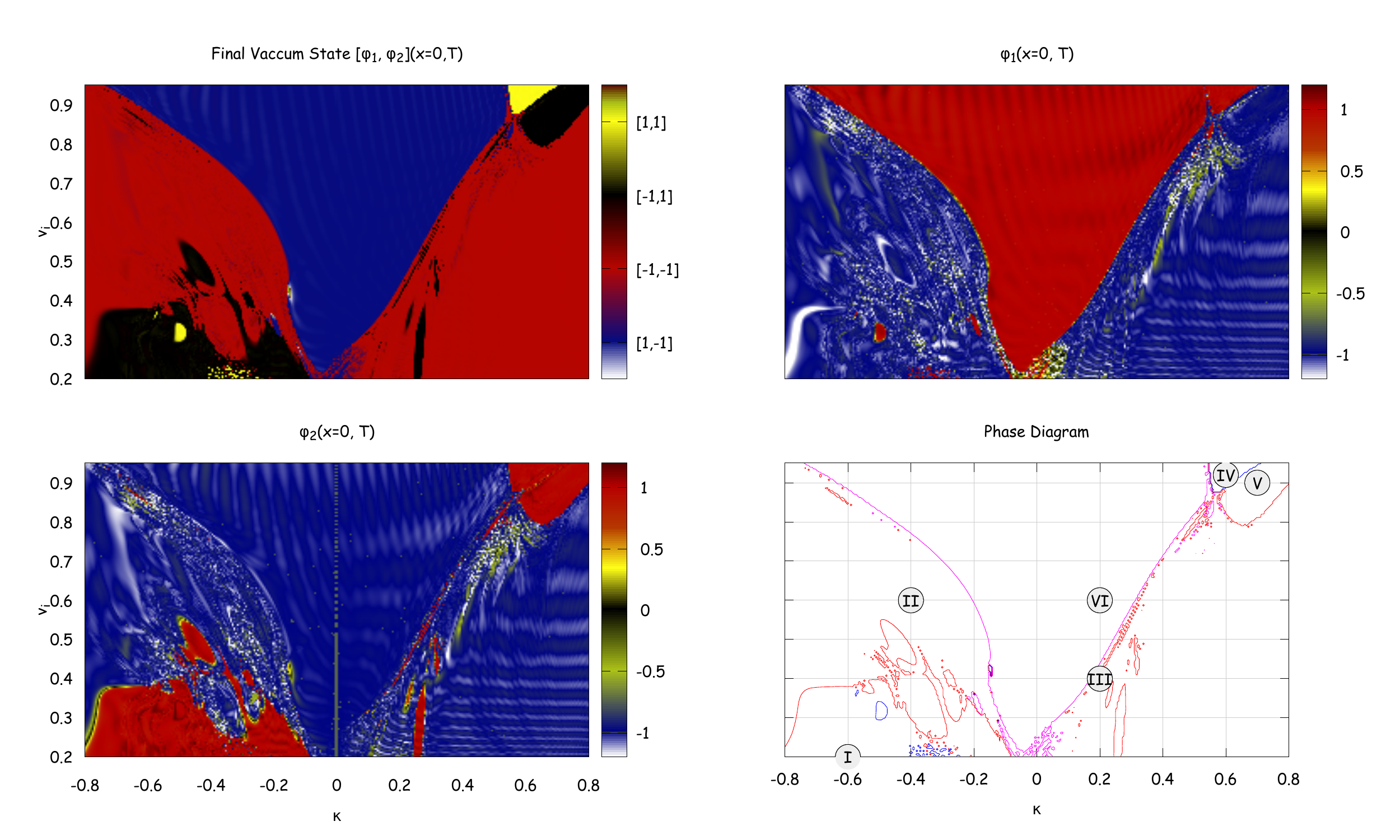}
\vspace{0.2cm}
\caption{\small (Color online) Final state of the double kink with a lump kink collision $(1,1)+(-1,0)$.
}\label{fig:DoublePhaseDiag3}
\end{figure}

\begin{figure}
\centering
\includegraphics[width=0.8\linewidth,angle=0]{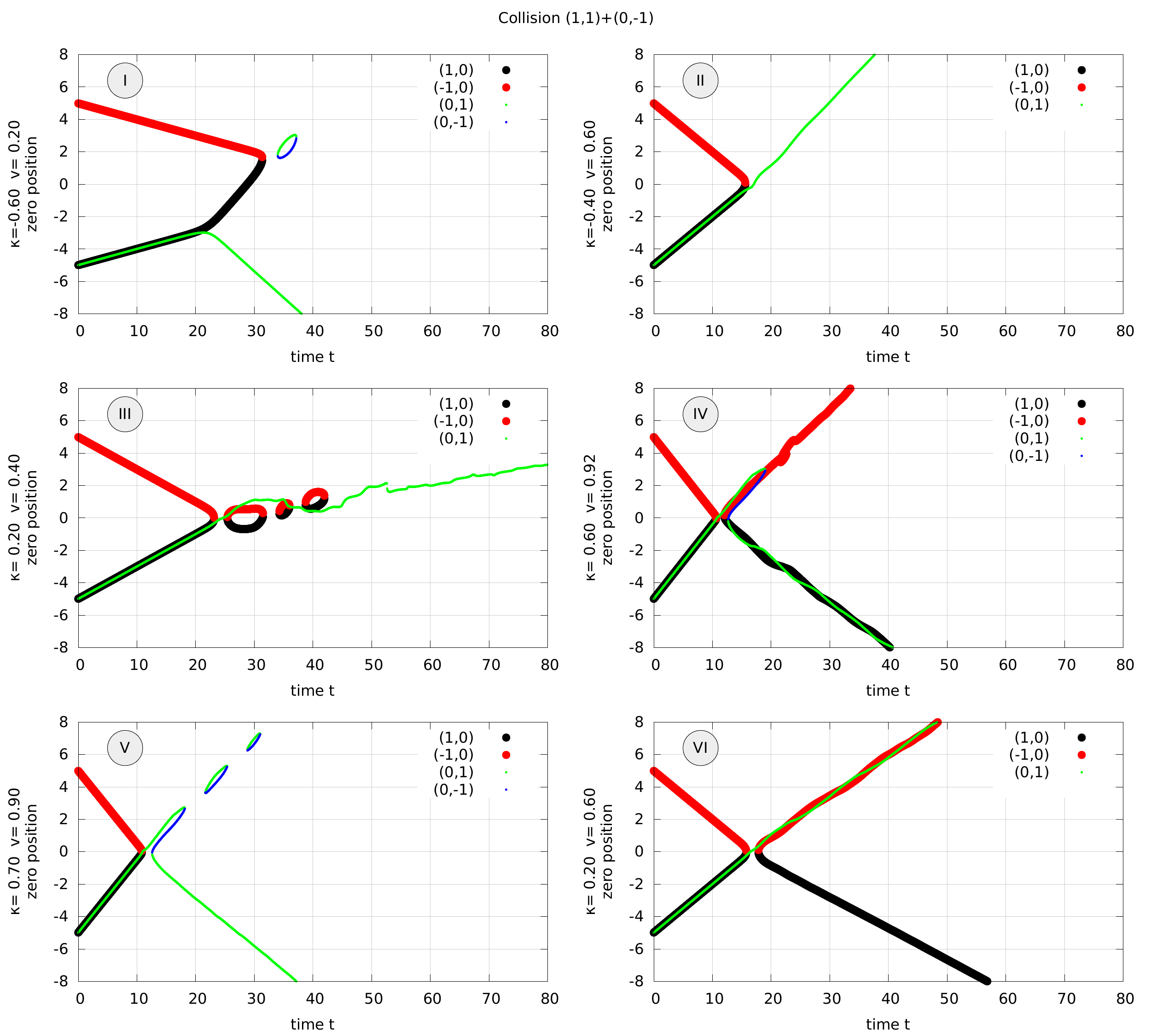}
\vspace{0.2cm}
\caption{\small (Color online) Example of topological zeros motion for collision of a double kink with a lump kink $(1,1)+(-1,0)$ with final states
(I) unstable double kink decays to pair of lump kinks, one is ejected and
second annihilates with the other kink $(1,1)+(-1,0)\to (0,1)+(1,0)+(-1,0)\to(0,1)+(0,0)$,
(II) annihilation in the first sector and passage in the second sector,
(III) annihilation in the first sector, the created oscillon is coupled to the remaining kink in the
second sector $(0,0)+(0,1)$,
(IV) bounce $(1,1)+(-1,0)$,
(V) annihilation in the first sector and bounce in the second sector $(0,1)+(0,0)$,
(VI) bounce in one sector and passage in the second sector $(1,0)+(-1,1)$.
}\label{fig:DoubleZeros3}
\end{figure}

\section{Spectral structure of linear perturbations}
There is certain similarity between chaotic structures observed in the collision of the kinks in the two-component
model \re{Lag} and $K\bar K$ resonance bouncing in the usual $\phi^4$ model \cite{Campbell:1983xu,Anninos:1991un}.
This analogy can be evidently seen in the collision of the lump kinks, however the chaotic structure of the collision between the
double kink and the lump kink is much more complicated.

To investigate this analogy in more details we have to make use of the linear expansion around the solitons of the model
under consideration assuming perturbations of both components
$(\phi_1 + \eta_1e^{i\omega t}, \phi_2+\eta_2e^{i\omega t})$, here the unperturbed configuration $(\phi_1, \phi_2)$
corresponds to one of the solutions of the model \re{Lag}, the double kink or the lump kink.

Linearization of the system \re{eq:static} yields the matrix eigenvalues equation
\begin{equation} \label{fluct}
   \begin{bmatrix}
      D_1^2& -4\kappa \phi_1\phi_2\\
      -4\kappa \phi_1\phi_2& D_2^2\\
   \end{bmatrix}
\begin{bmatrix}
      \eta_1\\\eta_2
   \end{bmatrix}=
  \omega^2\begin{bmatrix}
      \eta_1\\\eta_2
   \end{bmatrix}
\end{equation}
where the operators $D_1^2, D_2^2$ are
\begin{subequations}
\begin{align}
   D_1^2 &= -\frac{d^2}{dx^2}+6\phi_1^2-2-2\kappa\phi_2^2,\\
   D_1^2 &= -\frac{d^2}{dx^2}+6\phi_2^2-2-2\kappa\phi_1^2.
   \end{align}
\end{subequations}
Note that in this section we
do not consider the coupling constant $\kappa$ as a perturbation parameter.

\subsection{Spectral structure of double kink}
First, let us consider the spectrum of linear oscillations around the double kink solution. In this case
$\phi_2 = \pm \phi_1 = \pm \frac{\tanh x}{\sqrt{1-\kappa}}$ and it is possible to diagonalise the system \re{fluct}
introducing new functions
\be \label{diag}
\xi_1 = \frac{1}{2}\left(\eta_1 + \eta_2 \right); \qquad \xi_2 = \frac{1}{2}\left(\eta_1 - \eta_2 \right).
\ee
Here the variable $\xi_1$ describes the motion of the center of mass of the system whereas the variable $\xi_2$
corresponds to the oscillations of the components about it.
Then the system of equations \re{fluct} is decoupled and we get
\begin{subequations}
\begin{align}
 \left(-\frac{d^2}{dx^2}+6\tanh^2 x-2\right)\xi_1&=\omega^2\xi_1,\\
 \left(-\frac{d^2}{dx^2}+\frac{6+2\kappa}{1-\kappa}\tanh^2x-2\right)\xi_2&=\omega^2\xi_2 .
 \label{eq:PTeller}
\end{align}
\end{subequations}
Evidently, the first of these equations coincides with the usual modified Peashle-Teller equation
that describes the linear excitations around the $\phi^4$ kink, it is well known the spectrum of fluctuations in this
case has one translational zero mode $\omega = 0$, the internal mode of the kink $\omega^2 = 3$,
which corresponds to the oscillations of the width of the kink, and the continuum modes $\omega^2= k^2+4$:
\begin{eqnarray}                                \label{modes}
\xi_1^{(0)} = \frac{1}{\cosh^2~x};\qquad
\xi_1^{(1)} = \frac{\sinh x}{\cosh^2~x};
\nonumber\\[4pt]
\xi_1^{(k)} = e^{ikx} (3 \tanh ^2~x - 3ik \tanh x -1 - k^2).
\end{eqnarray}
The second linearly independent solution is the complex conjugation of the above.

In the case under consideration, for example the excitation of the translational mode $\xi_1^{(0)}
= \frac{1}{2}\left(\eta_1^{(0)} + \eta_2^{(0)} \right)$ corresponds to the synchronous translation
of the double kink whereas the excitation of the internal mode $\xi_1^{(1)}$ corresponds to the
synchronous wobbling of both components.

The meaning of the second mode is different.
Indeed, the second equation (\ref{eq:PTeller}) is of the same type,
so it can be solved by reducing it to the hypergeometric
form (see, e.g. \cite{KST}). Here, however, we shall follow another route. Let us introduce the ladder operators
\be \label{ladder}
\hat a^\dagger=-\frac{d}{dx}+n\tanh x; \qquad
 \hat a=\frac{d}{dx}+n\tanh x,
\ee
which allows us to transform the equation \re{eq:PTeller} to the form
\be \label{degree:n}
   \left(\hat a^\dagger \hat a+n-2\right)\xi_2=\omega^2\xi_2
\ee
Here we introduced a new parameter
\begin{equation}
    n = -\frac{1}{2}+\frac{1}{2}\sqrt{\frac{25+7\kappa}{1-\kappa}} \, .
\end{equation} 
which is defined as a solution of the algebraic equation
\begin{equation}
n(n+1)=\frac{6+2\kappa}{1-\kappa} \, .
\end{equation}

\begin{figure}
\centering
\includegraphics[width=0.8\linewidth,angle=0]{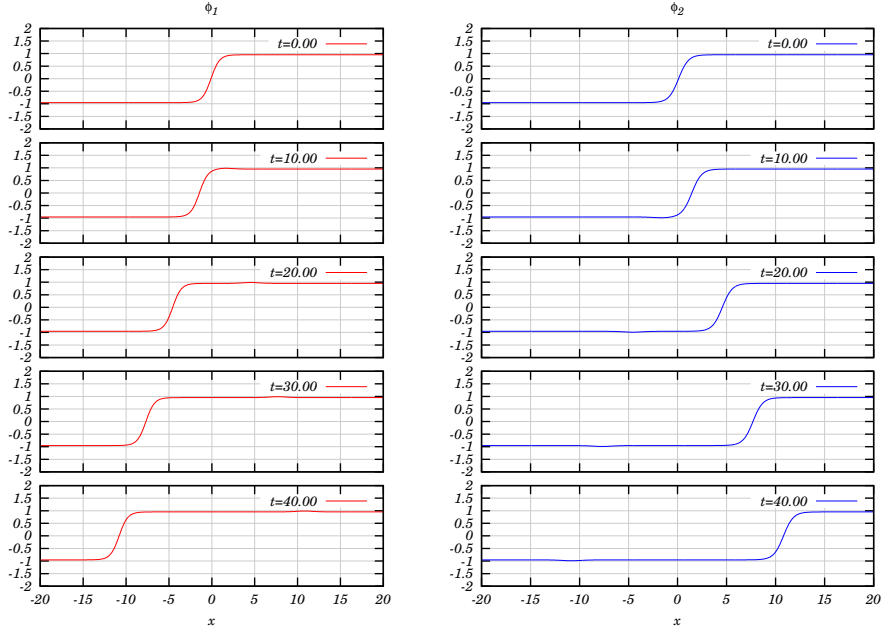}
\caption{\small (Color online) Decay of the double kink  at $\kappa=-0.1$.
Initial configuration is taken as $\phi_1(x,0) = u\tanh(x+0.1)$ and $\phi_2(x,0) = u\tanh(x-0.1)$. Excitation
of the lowest translational mode $\xi_2^{(0)}$ leads to decay of the double kink into
two lump kinks moving in opposite directions.
}\label{fig:8}
\end{figure}
Evidently, the ground state is defined as the state annihilated by the operator $\hat a$, i.e.,
\begin{equation}
   \hat a \xi_2=0.
\end{equation}
Thus, up to a normalisation factor, the solution to this equation is
\begin{equation}\label{eq:ground}
   \xi_2^{(0)}=\frac{1}{\cosh^n\!x}\, .
\end{equation}
Here the integer $n$ labels the internal modes of the double kink system, their number increases with coupling constant
since
\begin{equation}
   \kappa(n) = \frac{n^2+n-6}{n^2+n+2}.
\end{equation}
Thus, for $-1<\kappa<0$ there is only one bound state, for $0<\kappa<3/7$ there are two internal modes,
for $3/7<\kappa<7/11$ there are three internal modes, etc. Indeed, the properties of the modes are specified
by the potential $U(x)=n(n+1)\tanh^2x-2 $ of the equation (\ref{eq:PTeller}), it is getting deeper as the integer $n$
increases. The number of bound states trapped by this potential then also increases.

Note that the frequency of the lowest mode is $\omega^2 = n-2$, thus if $n<2$, or equivalently,
if $\kappa < 0$, there is a negative mode
in the spectrum and the lump kink becomes unstable with respect to perturbations. If the coupling constant $\kappa$ remains
positive, the lump kink is stable and excitation of this mode $\xi_2^{(0)}
= \frac{1}{2}\left(\eta_1^{(0)} - \eta_2^{(0)} \right)$ corresponds to the oscillations of the components of the double
kink about the center of mass of the system.

Our numerical studies confirmed this conclusion. An example of the decay of the double kink at $\kappa = -0.1$
is displayed in Fig.~\ref{fig:8}, we can see that the excitation
of the lowest translational mode $\xi_2^{(0)}$ leads to decay of the double kink into two lump kinks which
go off in contrary directions. On  the other hand, the same perturbative distortion of the double kink at positive
coupling $\kappa = 0.5$ and consequent excitation of the wobbling collective mode does not destroy the configuration.

The  vibrational mode of the double kink  has the following form (up to a normalisation factor):
\begin{equation}
    \xi_2^{(1)} = \frac{\sinh\,x}{\cosh^nx}.
\end{equation}
The corresponding frequency depends on integer $n$ as  $\omega = \sqrt{3(n-1)}$.
Note, that action of the creation operator on the ground state (\ref{eq:ground}) yields:
\begin{equation}
   \hat a^\dagger\xi_2^{(0)}=\frac{2n\sinh x}{\cosh^{n+1}x}.
\end{equation}
Evidently, this is a $(n+1)$-th bound state. So in order to obtain the $n$-th bound
state, the creation operator should act on the $(n-1)$-th mode.
Indeed, the algebra of the ladder operators  \re{ladder} is
\begin{equation}
   [\hat a,\hat a^\dagger] = \frac{2n} {\cosh^2x} \, .
\end{equation}
Therefore
\begin{equation}
\left(\hat a^\dagger \hat a+n-2\right)\hat a^\dagger =
\hat a^\dagger\left(\hat a^\dagger \hat a+n-2+2n-2n\tanh^2\,x\right)=
\hat a^\dagger \left(\hat {\tilde a}^\dagger \hat {\tilde a}+ n-2+p\right),
\end{equation}
where $p=2n+1$ and we introduce  new ladder operators of degree $(n-1)$ (cf. Eq. \re{ladder})
\begin{equation}
\hat{\tilde a}^\dagger = -\frac{d}{dx}+(n-1)\tanh x \, , \qquad
 \hat{\tilde a} = \frac{d}{dx}+(n-1)\tanh x \, .
\end{equation}
This, while the function $\xi_2^{(k)}$ is an eigenfunction of the operator of degree $n$ \re{degree:n} and the corresponding
eigenvalue is $\omega^2$, the action of the raising operator $\hat{\tilde a}$ transforms it to the eigenfunction  of
the operator of degree $(n-1)$ with eigenvalue $\omega^2+p$.

\subsection{Spectral structure of lump kink}
Investigation of the spectrum of the linear lump kinks fluctuations is a bit more complicated problem because, unlike in
the double kink case, there is no analytical solutions for the classical solutions $\phi_1$ and $\phi_2$. Thus, our
analysis relies on numerical methods, we have to find
solutions to the system of equations \re{fluct} using shooting in two dimensions or applying the spectral Chebyshev
method.

\begin{figure}
\centering
\includegraphics[width=0.8\linewidth,angle=0]{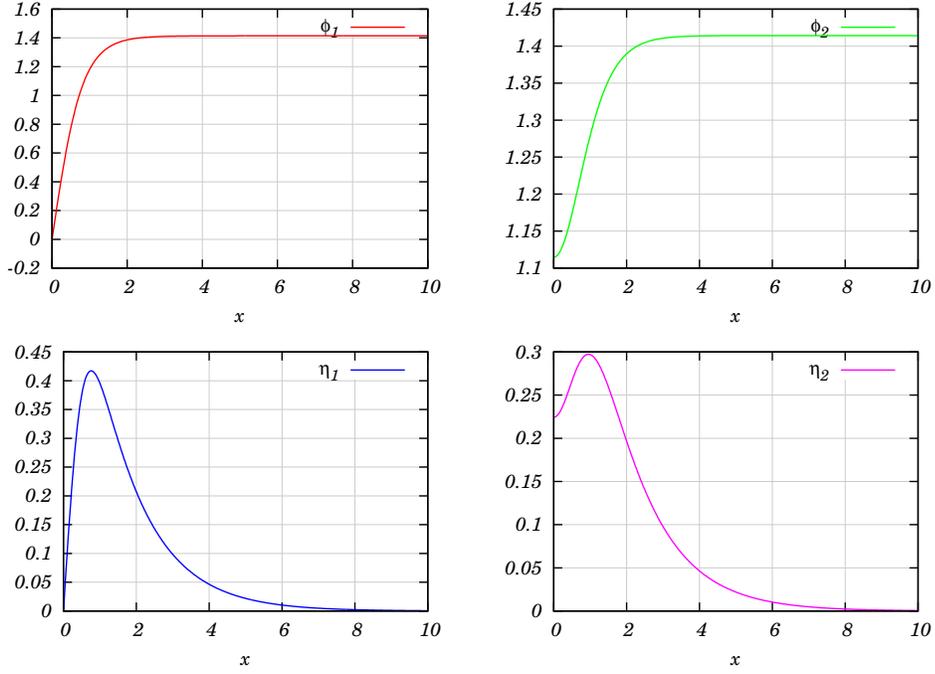}
\caption{\small Profiles of the lump kink components $\phi_1, \phi_2$
together with the profiles of the corresponding internal modes $\eta_1, \eta_2$ are presented at
$\kappa=0.5$, the frequency of the internal mode is
$\omega=1.8537$.
}\label{fig:9}
\end{figure}

\begin{figure}
\centering
\includegraphics[width=0.8\linewidth,angle=0]{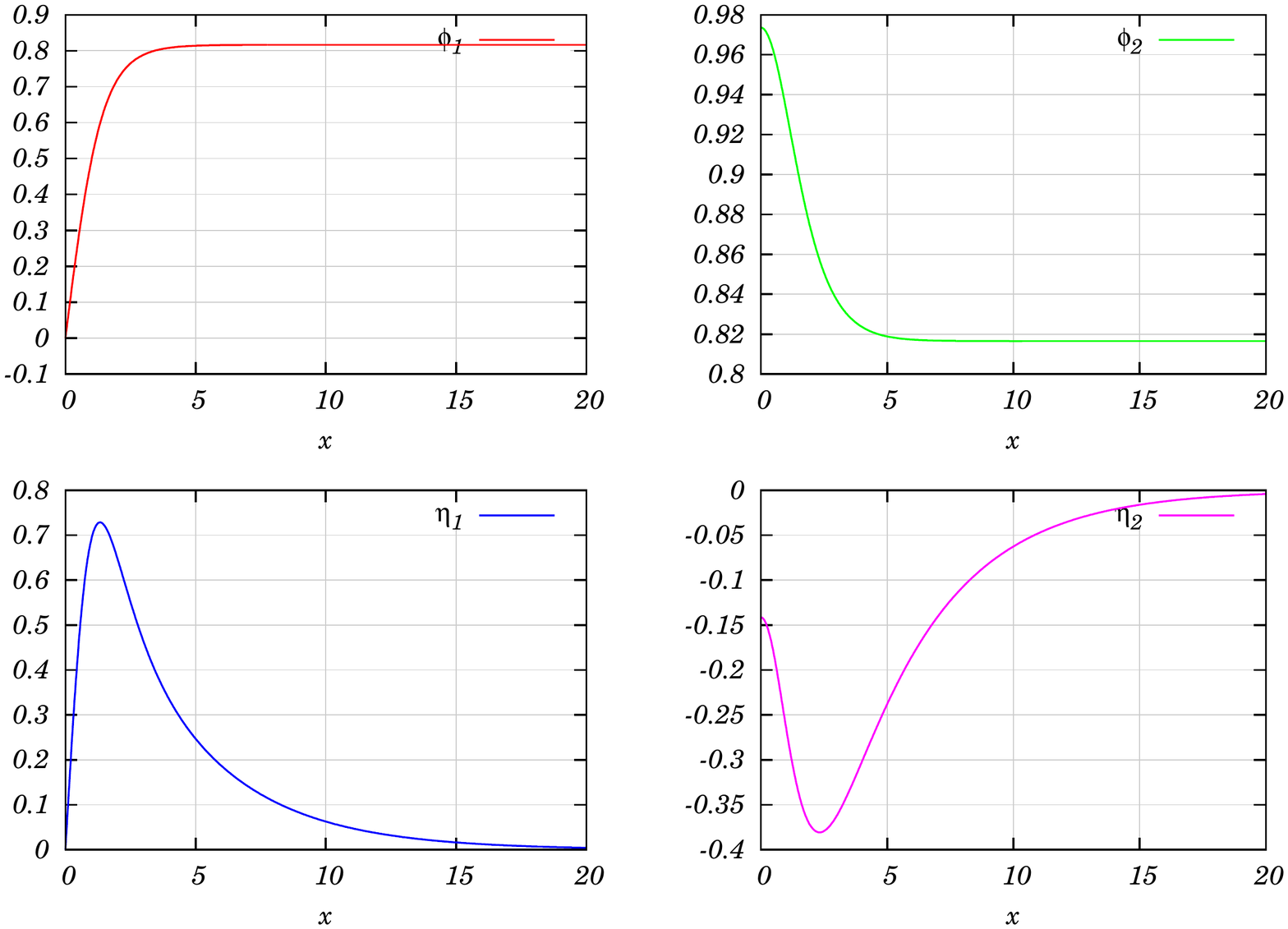}
\vspace{0.2cm}
\caption{\small Profiles of the lump kink components $\phi_1, \phi_2$
together with the profiles of the corresponding internal modes $\eta_1, \eta_2$ are presented at
$\kappa=-0.5$, the frequency of the internal mode is
$\omega=1.12245$.
}\label{fig:10}
\end{figure}

Note the the system of the linearized equations \re{fluct} can be diagonalised  on the spacial asymptotics
where the fields approach the vacuum value: $\phi_1 = \phi_2 = 1/\sqrt{1-\kappa}$ as $x\to \pm \infty$. Introducing the
same variables $\xi_1, \xi_2$ \re{diag}, we obtain then decoupled system of equations
\begin{subequations}
\begin{align}
 \left(-\frac{d^2}{dx^2}+4\right)\xi_1&=\omega^2\xi_1,\\
 \left(-\frac{d^2}{dx^2}+4~\frac{1+\kappa}{1-\kappa}\right)\xi_2&=\omega^2\xi_2.
\end{align}
\end{subequations}
Thus, there are 2 different excitations about the vacuum in our model \re{Lag} with 2 different masses
$m_1=2$ and $m_2=2\sqrt{\frac{1+\kappa}{1-\kappa}}$, they are degenerated only in the special decoupled case $\kappa=0$.
One of the interesting manifestation of the presence of two different excitations in the spectrum of linear
fluctuations in our model is related with behavior of the lump kink under the influence of an incident wave.
It is known in the usual one-component $\phi^4$ theory the kink starts to accelerate in the direction
of the incoming wave \cite{Rom:2004,Forgach} with acceleration proportional to the 4-th power of amplitude of incident wave. 
This effect is known as the negative radiation pressure. 
In our model $|\phi_1|=|\phi_2|$ symmetry restores exactly the $\phi^4$ model so within this sector of theory the same phenomenon takes place.

However in our system there is also a different possibility. Because of the difference in masses of small perturbation around the vacuum another mechanism of the negative radiation pressure (proportional to the square of the amplitude of the wave) can be observed \cite{Romanczukiewicz:2008hi}.
In the case under consideration numerical simulations confirm that if the frequency of the incoming wave
is $\omega^2 = m_1^2 + k^2$ the kink accelerates towards the source of the radiation, however if the frequency of
the incoming wave is $\omega^2 = m_2^2 + k^2$ the result is inverted, i.e. the
positive radiation pressure is observed.

\begin{figure}
\centering
\includegraphics[width=0.6\linewidth,angle=0]{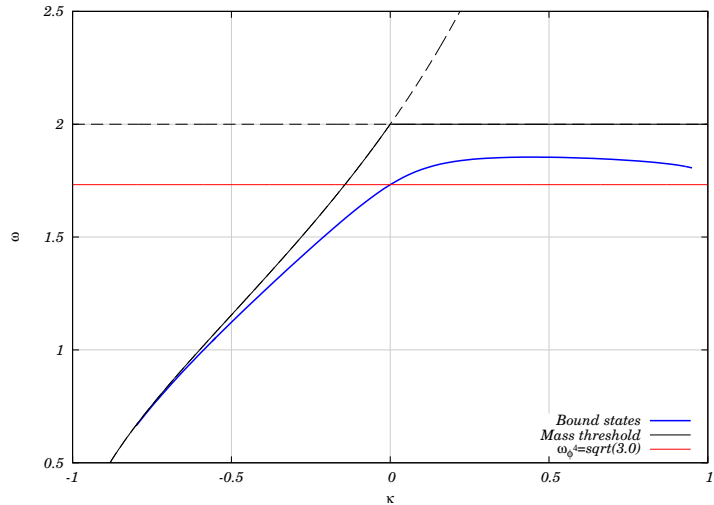}
\vspace{0.2cm}
\caption{\small Frequency of the internal mode of the lump kink is presented as a
function of coupling constant $\kappa$.
}\label{fig:11}
\end{figure}

Thus, the masses of the bound states of the lump kink are restricted by the lowest of the asymptotic masses
$m_1$ and $m_2$. We found the solution of the eigenvalue problem numerically using the spectral Chebyshev
method. The results are presented in Figs.~\ref{fig:9},\ref{fig:10}. Evidently, the internal mode of
the kink component is antisymmetric while the internal mode of the lump component is symmetric. Considering
the dependency of the frequency of the internal mode on the coupling constant $\kappa$ we found that there is no
second internal mode in the system, the second mass threshold is decreasing faster than the
frequency of the oscillation mode (cf. Fig.~\ref{fig:11}).

%%%%%%%%%%%%%%%%%%%%%%%%%%%%%%%%%%%%%%%%%%%%%%%%%%%%%%%%%%%%%%%%%%%%%%%%%%%%%%
\section{Conclusions}
%%%%%%%%%%%%%%%%%%%%%%%%%%%%%%%%%%%%%%%%%%%%%%%%%%%%%%%%%%%%%%%%%%%%%%%%%%%%%
Motivated by the recent interest in investigation of the remarkable resonance
structures in the parameter space of the kink-antikink system,
we have studied dynamical properties of the soliton solution of the two-component coupled $\phi^4$ system.
There are two different types of the solitons, the double kink and the lump kink,
in both cases our numerical simulations of the collisions between the various types of the solitons reveal some
chaotic resonance behavior which  however,
is rather different from the usual pattern of the
$K\bar K$ collision in the one-component $\phi^4$ model.

Considering collision of two lump kinks in the same sector below some critical velocity
we observed the resonance energy exchange
between the sectors with a sequences of bouncing windows similar with the quasi-fractal dynamics observed
in the usual $\phi^4$
model. However the structure of the resonance windows in the former case is not so regular,
also some false windows appear. At the collisions of the lump kinks above the critical velocity we observed
flip of the sectors and consequent scattering of the solitons. In the ultrarelativistic limit another process starts to
dominate, the collision of the kinks in the first sector is accompanying by the collision of the lumps in the second sector
which produces kink-antikink pairs.

We also analysed the collision of the lump kinks in the different sectors and the most complicated process of the douple kink -
lump kink collision. In the former case the resonance structure is very regular, it is qualitatively similar to the structure
which appears in the one-component model.
However the double kink - lump kink collision may lead to various results, there
are different channels of this process,
the most interesting case is related with collisions at relatively high impact velocities, then
a new type of self-similar fractal structure appears. These results need more rigorous investigation which is currently
in progress, in particular it would be interesting to find an effective theory which could explain such a behavior.
Evidently such a complicated behavior is
related with the spectrum of excitations which may be excited in the collision of the solitons and
affect the mechanism of the energy exchange between the solitons.

Investigation of the linear stability of the soliton solutions showed that for certain range of values of the coupling constant,
the double kinks are unstable with respect to linear perturbations,
they decay into pair of two lump kinks.

We also investigated the spectrum of perturbations of the soliton solutions on the two-component model.
We found that in the case of the double kink configuration, in addition to the expectable counterparts of the translational and
internal modes, there is a tower of bound states whose number depends on the strength of the coupling between the sectors. Evidently,
excitation of these states in the process of the collision of the kinks will strongly affect the mechanism of the energy transfer.
Considering the spectrum of linear fluctuation about the lump kinks we found there are excitations of 2 different masses, however
there is only one internal mode in the system for all range of values of the coupling constant.

It remains to systematically analyze the effect of interaction of the soliton solutions of the two-component model
with an incoming wave in perturbation theory, it will answer the question if the effect of negative radiation pressure is also
presented in this case. 

As a direction for future work, it would be interesting
to construct an effective collective coordinate Lagrangian for the two-component model which will capture the most
important degrees of freedom in the soliton collision.

We thank Patrick Dorey and Wojtek Zakrzewski  for enlightening discussions.
Ya.S.~is very grateful to Stephane Nonnenmacher for kind hospitality at the SPhT, CEA Saclay where part of this work
was done. A.H.~gratefully acknowledges support from the organizers of the 52. Cracow School of Theoretical Physics. 
This work is supported by the A.~von Humboldt Foundation (Ya.S.).

%%%%%%%%%%%%%%%%%%%%%%%%%%%%%%%%%%%%%%%%%%%%%%%%%%%%%%%%%%%%%%%%%%
%\begin{small}

%\end{small}
% 
% \newpage
% \appendix
% \section{Corrections}
% \subsection{Already made}
% \begin{itemize}
%    \item Small typos and some minor changes throughout the text (marked by the new \verb|\alternative| and \verb|\corrected| commands)
%    \item Section on the projection method was changed.
%    \item fft of the created double kink was added
%    \item subsections for different types of collisions
% \end{itemize}
% 
% \subsection{To do}
% \begin{itemize}
%    \item Ledder operators to the Apendix? -- stays as it is
%    \item More notes on the radiation pressure.-- to the next paper
%    \item some words on the phase diagrams? -- include all the phase diagrams.
% \end{itemize}

\end{document}